\documentclass[sn-mathphys,Numbered]{sn-jnl}

\usepackage{graphicx}%
\usepackage{multirow}%
\usepackage{amsmath,amssymb,amsfonts}%
\usepackage{amsthm}%
\usepackage{mathrsfs}%
\usepackage[title]{appendix}%
\usepackage{xcolor}%
\usepackage{textcomp}%
\usepackage{manyfoot}%
\usepackage{booktabs}%
\usepackage{algorithm}%
\usepackage{algorithmicx}%
\usepackage{algpseudocode}%
\usepackage{listings}%
\usepackage{bm}%

\theoremstyle{thmstyleone}%
\theoremstyle{thmstyletwo}%

\theoremstyle{thmstylethree}%

\newcommand{\be}{\begin{equation}}
	\newcommand{\ee}{\end{equation}}
\newcommand{\eq}[1]{(\ref{#1})}
\newcommand{\bit}{\begin{itemize}}  \newcommand{\eit}{\end{itemize}}

\def\bea{\begin{eqnarray}}
	\def\eea{\end{eqnarray}}

\def\a{\alpha}        \def\dA{{\dot A}}
\def\b{\beta}         
\def\g{\gamma}      
\def\d{\delta}      
\def\e{\epsilon}          
\def\f{\phi}          

\def\k{\kappa}  
\def\l{{\lambda}} \def\L{\Lambda}  
\def\m{\mu} \def\n{\nu}  
\def\o{\omega}  
\def\p{\pi}   
\def\r{\rho}  
\def\s{\sigma}

\def\cA{{\cal A}}  \def\cC{{\cal C}}  
\def\cD{{\cal D}}  \def\cF{{\cal F}}  
 \def\cH{{\cal H}} \def\cI{{\cal I}}  
\def\cJ{{\cal J}}  \def\cL{{\cal L}}

 \def\cT{{\cal T}}

\def\ot{\tilde{\Omega}} 

\def\pa{\partial}   

\def\ove#1{\frac{1}{#1}}

\newcommand{\lrbrk}[1]{\left(#1\right)}

\def\w{{\wedge}}

\renewcommand{\vec}[1]{\bm{#1}}

\begin{document}

\title[On sufficient conditions for degrees of freedom counting of
multi-field generalised Proca theories]{On sufficient conditions for degrees of freedom counting of
	multi-field generalised Proca theories\footnote[2]{Matched with version submitted to General Relativity and Gravitation. Title is changed to match with published version. This preprint has not undergone peer review or any post-submission improvements or corrections. The Version of Record of this article is published in General Relativity and Gravitation, and is available online at https://doi.org/10.1007/s10714-023-03191-8.}}


\author[1,2]{\fnm{Sujiphat} \sur{Janaun}}\email{sujiphatj62@nu.ac.th}

\author*[2]{\fnm{Pichet} \sur{Vanichchapongjaroen}}\email{pichetv@nu.ac.th}

\affil[1]{\orgdiv{Department of Physics, Faculty of Science,
		Naresuan University}, \orgaddress{Phitsanulok, \postcode{65000}, \country{Thailand}}}

\affil[2]{\orgdiv{The Institute for Fundamental Study ``The Tah Poe Academia Institute'', 
		Naresuan University}, \orgaddress{Phitsanulok, \postcode{65000}, \country{Thailand}}}

\abstract{We derive sufficient conditions for theories consisting of multiple vector fields, which could also couple to external fields, to be multi-field generalised Proca theories. The conditions are derived by demanding that the theories have the required structure of constraints, giving the correct number of degrees of freedom. The Faddeev-Jackiw constraint analysis is used and is cross-checked by Lagrangian constraint analysis. To ensure the theory is constraint, we impose a standard special form of Hessian matrix. The derivation benefits from the realisation that the theories are diffeomorphism invariance (or, in the case of flat spacetime, invariant under Lorentz isometry). The sufficient conditions obtained include a refinement of secondary-constraint enforcing relations derived previously in literature, as well as a condition which ensures that the iteration process of constraint analysis terminates. Some examples of theories are analysed to show whether they satisfy the sufficient conditions. Most notably, due to the obtained refinement on some of the conditions, some theories which are previously interpreted as being undesirable are in fact legitimate, and vice versa. This in turn affects the previous interpretations of cosmological implications which should later be reinvestigated.}




\maketitle

\section{Introduction}\label{sec1}

Physical phenomena are well described fundamentally by field theories, which provide fundamental laws and mechanisms from which phenomena arise. Many important phenomena can be described by theories of vector fields. For example, light can be described as a massless particle arising from the quantisation of a vector gauge field.
Force carriers of weak interactions \cite{Glashow:1959wxa}, \cite{Salam:1959zz}, \cite{Weinberg:1967tq} are described by vector fields which gain mass due to spontaneous symmetry breaking mechanism \cite{Higgs:1964ia}, \cite{Higgs:1964pj}, \cite{Englert:1964et}, \cite{Guralnik:1964eu}. Theories describing this type of vector fields have gauge symmetry which is spontaneously broken in the vacuum. More recently, important physical phenomena especially in cosmology can also be described by another type of massive vector field theories, in which gauge symmetry is explicitly broken due to the presence of the explicit mass terms. Furthermore, these theories are considered as effective field theories. Attempts to describe cosmological phenomena, for example, primordial inflation \cite{Guth:1980zm} and late-time accelerated expansion \cite{SupernovaCosmologyProject:1998vns}, \cite{SupernovaSearchTeam:1998fmf} using theories of massive vector fields are given for example in \cite{Rodriguez:2017wkg}, \cite{Gomez:2019tbj}, \cite{Garnica:2021fuu}.

In order to obtain a better understanding of vector-field-related phenomena, one of the important steps would be to classify vector theories and give the most general form of theories of each type. In particular, the criteria of the classification would be based on the constraint structure. For theories which describe a single vector field, the most useful types would be theories which generalise Maxwell theory and those which generalise Proca theory \cite{Proca:1936fbw}. The Dirac-Born-Infeld theory \cite{Born:1934gh}, \cite{Dirac:1962iy} is an example of theoy which generalises Maxwell theory. As for theories which generalise Proca theories, the notable constructions, with the aim to describe cosmological phenomena, start from the references \cite{Tasinato:2014eka} and \cite{Heisenberg:2014rta} (see \cite{Hull:2015uwa}, \cite{Allys:2015sht}, \cite{Jimenez:2016isa}, \cite{Allys:2016jaq}, \cite{Rodriguez:2017ckc} for developments; see also \cite{Heisenberg:2018vsk} for a review). 
The idea of \cite{Heisenberg:2014rta} is to impose the condition which we will call, for definiteness, ``the special Hessian condition'', in which time-time and time-space components of Hessian vanishes while determinant of space-space components is non-zero.
The reference \cite{Sanongkhun:2019ntn} shows by using constraint analysis that vector theories which are local, diffeomorphism invariance, having Lagrangian containing up to first order derivative in time (which ensures that the theory is free of Ostrogradski instability \cite{Ostrogradsky:1850fid}), as well as passing the special Hessian condition are likely to be generalised Proca theories.
Further generalisations to generalised Proca theories are possible, for example, references \cite{Heisenberg:2016eld}, \cite{GallegoCadavid:2019zke}, \cite{BeltranJimenez:2019wrd} construct beyond generalised Proca theories, references \cite{deRham:2021yhr}, \cite{deRham:2021efp}, \cite{ErrastiDiez:2022qvd}, \cite{deRham:2023brw} construct Proca-Nuevo\footnote{Note that \cite{deRham:2023brw} points out that \cite{ErrastiDiez:2022qvd} has obtained incorrect secondary constraint. So the result of \cite{ErrastiDiez:2022qvd} are not correct. We thank Claudia de Rham for letting us know this recent development and related discussions.}. 

When considering theories which
describe systems of multiple vector fields,
one would expect that they would simply be describable by several systems of single vector fields arbitrarily interacting with each other. It turns out, however, that the interactions cannot be arbitrary. Further conditions are required. As shown in \cite{ErrastiDiez:2019trb}, the special Hessian condition is not sufficient to ensure that theories legitimately contain only the required degrees of freedom. Further conditions called ``secondary-constraint enforcing relations'' should be imposed. Constructions of theories satisfying the special Hessian conditions and secondary-constraint enforcing relations are given in, for example, \cite{ErrastiDiez:2019ttn}, \cite{GallegoCadavid:2020dho}.
A more ambitious generalisation is provided by \cite{ErrastiDiez:2020dux} in which the systems of any fields, not necessarily vector fields, whose Lagrangian depends up to the first order derivative in the fields are attempted to be classified.

In principle, the formulation presented by \cite{ErrastiDiez:2019trb}, \cite{ErrastiDiez:2019ttn}, \cite{ErrastiDiez:2020dux} still needs small refinements. By nature of constraint analysis, including Lagrangian constraint analysis, time and space are not put on equal footing. Therefore, the analysis is carried out in the way that diffeomorphism invariance is not manifest in most steps (intermediate equations are usually not in the form where spacetime indices are contracted). This should be compensated by making use of the conditions that we will call ``diffeomorphism invariance requirements'', which are conditions automatically satisfied by any theory which is diffeomorphism invariant. Although trivial for each specific theory, these conditions are helpful for the simplifications of equations in intermediate steps of constraint analysis. Although these requirements are not used in \cite{ErrastiDiez:2019trb}, \cite{ErrastiDiez:2019ttn}, \cite{ErrastiDiez:2020dux}, we expect that they are crucial
in providing and simplifying sufficient conditions for theories to have the desired number of degrees of freedom. This is in fact demonstrated \cite{Sanongkhun:2019ntn} in the case of single-field generalised Proca theories. We will also demonstrate in our paper in the case of multi-field generalised Proca theories.

In fact, as will be discussed later in this paper, the conditions imposed by \cite{ErrastiDiez:2019trb}, \cite{ErrastiDiez:2019ttn} to ensure that the theories have secondary constraints are incorrect. There is one term missing from each of these conditions. Generically, this leads to incorrect counting of the number of degrees of freedom. Some theories which are previously interpreted as being undesirable in fact have the desired number of degrees of freedom, and vice versa. In principle, this could consequently lead to incomplete or even incorrect cosmological implications related to multi-field generalised Proca theories.

The goal of this paper is to derive sufficient conditions for theories to be multi-field generalised Proca theories. This is done by using Faddeev-Jackiw constraint analysis \cite{Faddeev:1988qp}, \cite{Jackiw:1993in}, \cite{Barcelos-Neto:1991rxi}, \cite{Barcelos-Neto:1991dhe} with the help of diffeomorphism invariance requirements. The steps to obtain the sufficient conditions are as follow. We first impose the special Hessian condition. This ensures that the theories are constraint as well as giving $n$ primary constraints where $n$ is the number of vector fields in the system. Next, extra conditions should be imposed \cite{ErrastiDiez:2019trb}, \cite{ErrastiDiez:2019ttn} which ensure that the theories have secondary constraints. The conditions we find actually give the correction to their counterpart obtained in \cite{ErrastiDiez:2019trb}, \cite{ErrastiDiez:2019ttn}. Further conditions should also be imposed to ensure that the symplectic two-form at the second iteration does not have a zero mode. If a theory passes all these requirements, then it is a multi-field generalised Proca theory. 

This paper is organised as follows. In section \ref{sec:analysis}, we consider theories of multiple vector fields which could also couple to external fields. We only consider the theories whose Lagrangians are local, diffeomorphism invariance, depend up to first order derivative of the vector fields and satisfy the special Hessian condition. We then proceed to use Faddeev-Jackiw constraint analysis on these theories and obtain the sufficient conditions for the vector sector to have the expected constraints structure and hence the correct number of degrees of freedom. We then make a cross-check in secion \ref{sec:checks} by using Lagrangian constraint analysis, which give rise to conditions which after transforming to phase space agree with those obtained in section \ref{sec:analysis}.
In section \ref{sec:application}, we discuss how to apply the sufficient conditions. In particular, we demonstrate in subsection \ref{subsec:samples} the usage of these conditions to check example theories previously presented in the literature. Most notably, we provide an example legitimate theory which is previously misinterpreted in the literature as containing extra degrees of freedom. We also provide an example undesirable theory which is previously misinterpreted in the literature as being legitimate. In subsection \ref{subsec:implications}, we argue how the reinterpretations given in subsection \ref{subsec:samples} would affect the study of cosmological implications previously presented in the literature. In section \ref{sec:conclusion}, we provide conclusion and discussion of results and possible future works.

\section{Analysis}\label{sec:analysis}

\subsection{Imposing special Hessian condition}
For definiteness, we consider theories in 4-dimensional spacetime. However, the analysis of this paper can easily be extended
to spacetime with other number of dimensions. We define Lagrangian density $\cL$ via
\be
S = \int d^4 x\ \cL.
\ee
We denote spacetime coordinates by $x^\m$ with $\m = 0,1,2,3.$ We also use other middle lower-case Greek indices $\m,\n,\r\in\{0,1,2,3\}$ to denote spacetime indices. We will denote spatial indices by using middle lower-case Latin indices $i,j,k,l\in\{1,2,3\}.$
When expressing field we will omit the dependence on time coordinate $t$.
We will also often drop the dependence on space variables $\vec{x}$ (but keep explicit other space variables e.g. $\vec{x}', \vec{y}, \vec{z}$). 
So for example $\varphi$ stands for $\varphi(t,\vec{x}),$ whereas $\varphi(\vec{y})$ stands for $\varphi(t,\vec{y}).$

We are interested in the class of multi-field generalised Proca theories which is a system of $n$ vector fields $A_\m^\a$ with $\a = 1,2,\cdots,n$ possibly
coupled to external fields, which might also include the metric $g_{\m\n},$ and their derivatives.
The external fields can be thought of as being dependent explicitly on time and space. For example, the system of multiple massive vector fields might be put in a flat or curved backgrounds and might also couple to other external fields. As for the notations, we use beginning lower-case Greek indices $\a,\b,\g\in\{1,2,\cdots,n\}$ to denote internal indices for vector fields. We call the external fields and their derivative collectively as $K.$

We consider theories whose Lagrangians are local, diffeomorphism invariance, free of Ostrogradski instability and depend up to first order derivatives of the vector fields. For definiteness, we call the space of the vector fields and their first order time derivatives as the tangent bundle.

In order for the vector sector to be constraint, the Hessian condition\footnote{The determinants in eq.\eq{Hessian-cond}, eq.\eq{Hssncon}, and eq.\eq{HssnconT} are defined as follows. We combine the two indices of each vector field into one collective index. The matrices appearing within the determinants then have two collective indices. Standard definition for determinant then applies.}
\be\label{Hessian-cond}
\det\lrbrk{\frac{\pa^2\cL}{\pa\dot A_\m^\a\pa\dot A_\n^\b}} = 0
\ee
should be satisfied. However, in this paper, we will restrict the study to theories satisfying condition
\be\label{Hssncon}
\frac{\pa^2\cL}{\pa\dA_0^{\a}\pa\dA_{\m}^{\b}}=0,\quad\det\lrbrk{\frac{\pa^2\cL}{\pa\dA_i^{\a}\pa\dA_j^{\b}}}\neq 0,
\ee
which would imply the Hessian condition \eq{Hessian-cond}.
For definiteness, let us call eq.\eq{Hssncon} as ``the special Hessian condition''. This condition has also been imposed by 
many references for example \cite{ErrastiDiez:2019trb}, \cite{ErrastiDiez:2019ttn}, \cite{GallegoCadavid:2020dho}, \cite{Allys:2016kbq}, \cite{ErrastiDiez:2021ykk}, in order to construct multi-field generalised Proca theories.

By requiring $\pa^2\cL/\pa\dA_0^\a\pa\dA_0^\b=0$, we see that $\cL$ should be at most linear in $\dA_0^\a.$
Then by using the condition $\pa^2\cL/\pa\dA_0^\a\pa\dA_i^\b=0,$ we see that the coefficient of the linear term does not depend on $\dot A_i^\a.$
Then imposing $\det(\pa^2\cL/\pa\dA_i^\a\pa\dA_j^\b)\neq 0$ exhausts all the requirements of eq.\eq{Hssncon}.

Therefore, theories we consider have Lagrangians of the form
\be\label{L-UV}
\cL
=T(A^\a_\m,\pa_i A_\m^\a,\dot A_i^\a, K)
+U_\b(A_\m^\a,\pa_i A_\m^\a,K)\dot{A}_0^\b,
\ee
subject to
\be\label{HssnconT}
\det\lrbrk{\frac{\pa^2T}{\pa\dA_i^\a\pa\dA_j^\b}} \neq 0.
\ee
Since these theories are diffeomorphism invariance, they satisfy conditions on $T, U_\b$ as given in Appendix \ref{app:diffeo}.
Further requirements will be imposed in order for the theory to possess the correct number of degrees of freedom. These requirements are known in the literature to allow secondary constraints and to terminate the process of constraint analysis \cite{ErrastiDiez:2019trb}, \cite{ErrastiDiez:2019ttn}, \cite{ErrastiDiez:2020dux}, \cite{ErrastiDiez:2021ykk}. 
The conditions which we will present are slightly differed from their counterparts in the literature. These differences, however, are important. Later in this section, we will comment on how and why they differ.

Euler-Lagrange equations for the vector fields are of the form
\be
\frac{\pa^2\cL}{\pa\dA_i^\a\pa\dA_j^\b}\ddot{A}_j^\b+\pa_j\frac{\pa\cL}{\pa\pa_jA_i^\a}-\frac{\pa\cL}{\pa A_i^\a} + \cdots=0,
\ee
\be
\dot{U}_\a+\pa_i\frac{\pa\cL}{\pa\pa_iA_0^\a}-\frac{\pa\cL}{\pa A_0^\a}=0,
\ee
where $\cdots$ are terms which do not contain $\ddot{A}^\a_\m$.
Since the Euler-Lagrange equations do not contain
time derivative with order higher than two,
the theories are free of Ostrogradski instability \cite{Ostrogradsky:1850fid} in the vector sector.
Furthermore, it is clear that the systems are free of Ostrogradski instability and are constrained as Euler-Lagrange equations are of second order derivative in time of $A_j^\b$ while there is only up to first order derivative in time for $A_0^\b.$
In section \ref{sec:checks}, we will start from these Euler-Lagrange equations and rederive, as a cross-check to the analysis of the present section, secondary-constraint enforcing relations \cite{ErrastiDiez:2019trb}, \cite{ErrastiDiez:2019ttn}. As to be seen in the analysis, the relations given in \cite{ErrastiDiez:2019trb}, \cite{ErrastiDiez:2019ttn} miss one term, which would invalidate some of their justifications on behaviour of example theories.

\subsection{Faddeev-Jackiw Constraint analysis}
We require that theories presented in the previous subsection
should have the correct number of degrees of freedom.
For this, we are going to make use of constraint analysis
using the Faddeev-Jackiw method \cite{Faddeev:1988qp}, \cite{Jackiw:1993in}, \cite{Barcelos-Neto:1991rxi}, \cite{Barcelos-Neto:1991dhe}. The analysis will give further conditions
that the theories should satisfy. We will use the notations and conventions similar to those used in \cite{Sanongkhun:2019ntn}, \cite{Toms:2015lza}.

\subsubsection{First iteration}
In order to transform from the tangent bundle to phase space,
one considers conjugate momenta.
Conjugate momenta for the Lagrangian eq.\eq{L-UV}
are
\be
\begin{split}
	\p_\b^\m
	&=\d_0^\m U_\b + \d_i^\m\frac{\pa T}{\pa\dot{A}_i^\b}.
\end{split}
\ee
These equations allow us to identify primary constraints
\be
\Omega_\b
=\p_\b^0 - U_\b.
\ee
The spatial components of conjugate momenta
are given by
\be\label{spatial-components-conjugate-momenta}
\p^i_\b
=\frac{\pa T}{\pa\dot{A}_i^\b}.
\ee
Because of the condition \eq{HssnconT},
these equations can be inverted to give
\be
\dot{A}_i^\b = \L_i^\b(A^\a_\m,\pa_i A_\m^\a,\p^i_\a,K).
\ee
Since we work in phase space, it would be convenient to define 
\be
\cT(A^\a_\m,\pa_i A_\m^\a,\Lambda_i^\a, K)
=T(A^\a_\m,\pa_i A_\m^\a,\dot{A}_i^\a, K)\bigg|_{\dot{A}_i^\a\to \Lambda_i^\a}.
\ee
Hamiltonian is given by
\be
\begin{split}
	\cH
	&=\p^\m_\a\dot{A}_\m^\a - \cL - \dot{\g}^\a\Omega_\a\\
	&\approx
	\p^i_\a\Lambda_i^\a
	-\cT - \dot{\g}^\a\Omega_\a,
\end{split}
\ee
where $\g^\a$ are Lagrange multipliers. Then let us start considering first iteration.
First order form of the Lagrangian is given by
\be
\cL_{FOF}
=\p^\m_\a\dot{A}_\m^\a
+\cL_v + \dot{\g}^\a\Omega_\a,
\ee
where
\be
\cL_v
\equiv
\cT
-\p^i_\a\Lambda_i^\a.
\ee
Symplectic variables are
\be
\xi^I
=(A_\m^\a,\p^\m_\a,\g^\a).
\ee
Canonical one-form is given by
\be
\cA
=\int d^3\vec{x}(\p^\m_\a\d A^\a_\m + \Omega_\a\d\g^\a).
\ee
So symplectic two-form is
\be
\begin{split}
	\cF
	&=\int d^3\vec{x}\bigg(\d\p^\m_\a\w\d A^\a_\m +
	\d\p_\a^0\w\d\g^\a - \frac{\pa U_\a}{\pa A_\m^\b}\d A_\m^\b\w\d\g^\a
	-\frac{\pa U_\a}{\pa \pa_i A_\m^\b}\d \pa_i A_\m^\b\w\d\g^\a\bigg).
\end{split}
\ee
Demanding $i_z\cF = 0$ gives
\be
z^{\p^\m_\a}
+ \frac{\pa U_\b}{\pa A_\m^\a} z^{\g^\b}
-\pa_i\lrbrk{\frac{\pa U_\b}{\pa \pa_i A_\m^\a} z^{\g^\b}}
=0,
\ee
\be
z^{A_\m^\a} + \d^0_\m z^{\g^\a} = 0,
\ee
\be
z^{\p^0_\a} - \frac{\pa U_\a}{\pa A_\m^\b}z^{A_\m^\b}
-\frac{\pa U_\a}{\pa \pa_i A_\m^\b}\pa_i z^{A_\m^\b}
= 0.
\ee
In order for these equations to be consistent,
the equation
\be\label{pre-ccc}
\begin{split}
	&
	\lrbrk{\frac{\pa U_\a}{\pa A_0^\b}
		-\frac{\pa U_\b}{\pa A_0^\a}
		+\pa_i\frac{\pa U_\b}{\pa \pa_i A_0^\a}} z^{\g^\b}
	\\
	&\qquad
	+\lrbrk{\frac{\pa U_\b}{\pa \pa_i A_0^\a}
		+\frac{\pa U_\a}{\pa \pa_i A_0^\b}}\pa_i z^{\g^\b}
	=0
\end{split}
\ee
has to be satisfied.
In fact as analysed in Appendix \ref{app:diffeo}
diffeomorphism invariance requires, among others, eq.\eq{diffeo-UpaiA0}.
So we are left with
\be\label{qz}
\lrbrk{\frac{\pa U_\a}{\pa A_0^\b}
	-\frac{\pa U_\b}{\pa A_0^\a}
	+\pa_i\frac{\pa U_\b}{\pa \pa_i A_0^\a}} z^{\g^\b}
=0.
\ee
Let us denote
\be
q_{\a\b}\equiv
\frac{\pa U_\a}{\pa A_0^\b}
-\frac{\pa U_\b}{\pa A_0^\a}
+\pa_i\frac{\pa U_\b}{\pa \pa_i A_0^\a}.
\ee
We are particularly interested in the case where $\textrm{rank}(q_{\a\b}) = 0$, that is
\be\label{ccc}
\frac{\pa U_\a}{\pa A_0^\b}
-\frac{\pa U_\b}{\pa A_0^\a}
+\pa_i\frac{\pa U_\b}{\pa \pa_i A_0^\a}
=0.
\ee
As will be seen later, enforcing these conditions would lead to $n$ secondary constraints. We are only interested in the class of theories with this constraint structure. This class include, for example, a theory of $n$ uncoupled generalised Proca fields (an analysis will be given in subsection \ref{subsec:samples}). On the other hand, if $\textrm{rank}(q_{\a\b})\neq 0,$ and we want the procedure not to terminate after the second iteration, the theory would either have undesired number of degrees of freedom or have first class constraints. Either of these cases are not what we are interested in.

As a cross-check, one may note that after imposing diffeomorphism invariance requirement,
\be
[\Omega_\a,\Omega_\b(\vec{x}')] \approx q_{\a\b}\d^{(3)}(\vec{x}-\vec{x}').
\ee
Therefore,
the condition eq.\eq{ccc} is equivalent to the vanishing of the Poisson's brackets of the primary constraints among themselves. That is
\be\label{ccc-pb}
[\Omega_\a,\Omega_\b(\vec{x}')]\approx 0.
\ee
In Dirac constraint analysis \cite{Dirac:1950pj}, \cite{dirac2001lectures}, if we demand the time evolution of primary constraints to vanish we would have, with $\cH_0 = \p^i_\a\L_i^\a - \cT,$
\be\label{tevo-prim}
\begin{split}
	&\int d^3\vec{x}'[\Omega_\a,\cH_0(\vec{x}')] + \int d^3\vec{x}'u^\b(\vec{x}')[\Omega_\a,\Omega_\b(\vec{x}')]	-\frac{\pa U_\a}{\pa K}\dot K\approx 0,
\end{split}
\ee
where it is understood that in the third term on LHS of eq.\eq{tevo-prim} there is a sum over the collection of the external fields and their derivatives.
If the conditions \eq{ccc-pb} are not fulfilled, i.e. $\textrm{rank}(q_{\a\b})\neq 0,$ eq.\eq{tevo-prim} would determine some components of $u^\b.$ So there will be less than $n$ secondary constraints. In the extreme case where $\textrm{rank}(q_{\a\b}) = n,$ i.e. $\det(q_{\a\b}) \neq 0,$
there is no secondary constraint. Furthermore, after classification, it is easy to see that all of these constraints are of second class. So the number of degrees of freedom is less than $3n,$ which is not desirable.

Note that in the tangent bundle, eq.\eq{ccc} can also be expressed as
\be\label{ccc-tb}
\frac{\pa^2 \cL}{\pa\dot A_0^\a\pa A_0^\b}
-\frac{\pa^2 \cL}{\pa A_0^\a\pa\dot A_0^\b}
+\pa_i\lrbrk{\frac{\pa^2 \cL}{\pa \pa_i A_0^\a\pa\dot A_0^\b}}
=0,
\ee
which is a correction to the secondary-constraint enforcing relations derived in \cite{ErrastiDiez:2019trb}, \cite{ErrastiDiez:2019ttn}.
Only the last term on the LHS of eq.\eq{ccc-tb} is not present in these references.
This could be due to the fact that their analysis discards the dependence on spatial derivatives of vector fields. While this is sufficient for the main purpose of counting the number of degrees of freedom, one should be careful with the conditions derived in the process. In order to make use of such conditions, one should appropriately restore the dependence on spatial derivatives of vector fields. It turns out that the restoration in this case is given by the inclusion of the third term on LHS of eq.\eq{ccc-tb}. As a consequence of the missing term in the secondary-constraint enforcing relations, behaviours of some theories receive incorrect interpretations. For example, 
a special case of theory presented in \cite{Allys:2016kbq} is interpreted by \cite{ErrastiDiez:2019trb} to contain extra degrees of freedom.
In fact, however, by a careful analysis to be discussed in subsection \ref{subsec:samples}, the theory is a legitimate multi-field generalised Proca theory since it has the desirable number of degrees of freedom.

Let us continue the Faddeev-Jackiw analysis. The zero mode of $\cF$ is
\be
\begin{split}
	z_1
	&=z^{\g^\a}\lrbrk{\frac{\d}{\d\g^\a} - \frac{\d}{\d A^\a_0}}
	+\lrbrk{-\frac{\pa U_\a}{\pa A_0^\b}z^{\g^\b}
		-\frac{\pa U_\a}{\pa \pa_i A_0^\b}\pa_i z^{\g^\b}}\frac{\d}{\d \p_\a^0}
	\\
	&\qquad
	+\lrbrk{-\frac{\pa U_\b}{\pa A_i^\a} z^{\g^\b}
		+\pa_j\lrbrk{\frac{\pa U_\b}{\pa \pa_j A_i^\a} z^{\g^\b}}}\frac{\d}{\d \p_\a^i},
\end{split}
\ee
subject to secondary-constraint enforcing relations \eq{ccc}.
Having obtained the zero mode, let us check whether there are further constraints in the system by considering
\be\label{further-constraints}
\begin{split}
	i_{z_1}\int d^3\vec{x}\ \d\cL_v
	&=\int d^3\vec{x}\bigg(-\frac{\pa\cT}{\pa A_0^\b} + \pa_i\frac{\pa\cT}{\pa\pa_i A_0^\b} 
	+\lrbrk{\frac{\pa U_\b}{\pa A_i^\a}
		+\frac{\pa U_\b}{\pa\pa_j A_i^\a}\pa_j}\L_i^\a\bigg)z^{\g^\b},
\end{split}
\ee
where we have used the identity
\be
\p^i_\a
=\frac{\pa\cT}{\pa \L_i^\a},
\ee
which is equivalent to eq.\eq{spatial-components-conjugate-momenta}.
The result from eq.\eq{further-constraints} gives secondary constraints
\be\label{tildeOmega}
\tilde{\Omega}_\b
=\frac{\pa\cT}{\pa A_0^\b} - \pa_i\frac{\pa\cT}{\pa\pa_i A_0^\b}
-\lrbrk{\frac{\pa U_\b}{\pa A_i^\a}
	+\frac{\pa U_\b}{\pa\pa_j A_i^\a}\pa_j}\L_i^\a-\frac{\pa U_\b}{\pa K}\dot{K},
\ee
which, written as functions,
\be
\tilde{\Omega}_\b
=\tilde{\Omega}_\b(A_\m^\a,\pa_i A_\m^\a,\pa_i\pa_j A_\m^\a, \p_i^\a,\pa_i\p_j^\a,K).
\ee
Note that when reading off the constraint \eq{tildeOmega}, there is also the contribution from external fields as presented in the last term on RHS. This is because the external fields are considered to be functions with explicit dependence on time. So when working out secondary constraints which essentially involves taking derivative of primary constraints with respect to time, the explicit time derivative of the external field should also be taken into account.

\subsubsection{Second iteration}\label{sec:second-it}
Having obtained new constraints from the first iteration, let us start the second iteration by including Lagrange multipliers corresponding to the new constraints. Symplectic variables are
\be
\xi^I
=(A_\m^\a,\p^\m_\a,\g^\a,\tilde{\g}^\a).
\ee
Canonical one-form is given by
\be
\cA
=\int d^3\vec{x}(\p^\m_\a\d A^\a_\m
+ \Omega_\a\d \g^\a
+ \tilde{\Omega}_\a\d\tilde{\g}^\a).
\ee
So symplectic two-form is
\be\label{cF-2nd-it}
\begin{split}
	\cF
	&=\int d^3\vec{x}\lrbrk{\d\p^\m_\a\w\d A^\a_\m
		+ \d \Omega_\a\w\d\g^\a
		+ \d\tilde{\Omega}_\a\w\d\tilde{\g}^\a}.
\end{split}
\ee
We may also denote the constraints and Lagrange multipliers as $\Omega^{(1)}_\a\equiv\Omega_\a, \Omega^{(2)}_\a\equiv\tilde{\Omega}_\a, \g_{(1)}^\a\equiv\g^\a, \g_{(2)}^\a\equiv\tilde{\g}^\a.$

When solving for zero mode of the symplectic two-form $\cF,$ equations involving Poisson's brackets would arise. In order to easily see this, it will be useful to define the notation for generalised derivatives $\pa_\cI$ as follows.
Suppose that $f$ and $g$ are functions of $A_\m^\a, \pa_i A_\m^\a, \pa_i\pa_j A_\m^\a,\cdots,$ $\p_\a^\m, \pa_i\p_\a^\m, \pa_i\pa_j \p_\a^\m,\cdots, K.$
So
\be
\begin{split}
	\frac{\d f}{\d A_\m^\a(\vec{z})}
	&=\frac{\pa f}{\pa A_\m^\a}\d^{(3)}(\vec{x}-\vec{z})
	+\frac{\pa f}{\pa\pa_i A_\m^\a}\pa_i\d^{(3)}(\vec{x}-\vec{z})
	+\frac{\pa f}{\pa\pa_i\pa_j A_\m^\a}\pa_i\pa_j\d^{(3)}(\vec{x}-\vec{z})
	+\cdots\\
	&\equiv
	\frac{\pa f}{\pa\pa_{\cI} A_\m^\a}\pa_\cI\d^{(3)}(\vec{x}-\vec{z}),
\end{split}
\ee
where summation over $\cI$ is understood.
Similarly,
\be
\begin{split}
	\frac{\d f}{\d \p_\a^\m(\vec{z})}
	&=\frac{\pa f}{\pa\pa_\cI\p_\a^\m}\pa_\cI\d^{(3)}(\vec{x}-\vec{z}).
\end{split}
\ee
Then in this notation Poisson's bracket can be written as
\be\label{PB-fg}
\begin{split}
	[f,g(\vec{y})]
	=&(-1)^{|\cJ|}\frac{\pa f}{\pa\pa_{\cI} A_\m^\a}\pa_\cI\pa_\cJ\!\!\lrbrk{\frac{\pa g}{\pa\pa_\cJ\p_\a^\m}\d^{(3)}(\vec{x}-\vec{y})}\\
	&
	-(-1)^{|\cJ|}\frac{\pa f}{\pa\pa_{\cI} \p_\a^\m}\pa_\cI\pa_\cJ\!\!\lrbrk{\frac{\pa g}{\pa\pa_\cJ A_\m^\a}\d^{(3)}(\vec{x}-\vec{y})},
\end{split}
\ee
where $|\cJ|$ is the order of partial derivatives of $\cJ,$ and summation over $\cI$ and $\cJ$ is understood.

Let us then find zero mode of $\cF.$
Demanding $i_z\cF = 0$ gives
\be\label{zpi-eq}
\begin{split}
	&
	z^{\p^\m_\b}
	- \sum_{\mathfrak{s}=1}^2(-1)^{|\cI|}\pa_\cI\lrbrk{z^{\g_{(\mathfrak{s})}^\a}\frac{\pa\Omega^{(\mathfrak{s})}_\a}{\pa\pa_\cI A_\m^\b}}
	= 0,
\end{split}
\ee
\be\label{za-eq}
\begin{split}
	&
	-z^{A^\b_\m}
	- \sum_{\mathfrak{s}=1}^2(-1)^{|\cI|}\pa_\cI\lrbrk{z^{\g_{(\mathfrak{s})}^\a}\frac{\pa\Omega^{(\mathfrak{s})}_\a}{\pa\pa_\cI \p^\m_\b}}
	=0,
\end{split}
\ee
\be
\pa_\cI z^{A_\m^\a}\frac{\pa\Omega^{(\mathfrak{s})}_\b}{\pa\pa_\cI A_\m^\a}
+\pa_\cI z^{\p^\m_\a}\frac{\pa\Omega^{(\mathfrak{s})}_\b}{\pa\pa_\cI \p^\m_\a}
=0,\quad \textrm{for }\mathfrak{s} = 1,2.
\ee
Eliminating $z^{A_\m^\a}$ and $z^{\p^\m_\a}$
and using the identity eq.\eq{PB-fg},
we obtain
\be\label{1st-result-eqFJ}
\sum_{{\mathfrak{s}}=1}^2\int d^3\vec{y} 
[\Omega^{(1)}_\a,\Omega^{(\mathfrak{s})}_\b(\vec{y})]z^{\g_{(\mathfrak{s})}^\b}(\vec{y})
=0,
\ee
\be\label{2nd-result-eqFJ}
\sum_{{\mathfrak{s}}=1}^2\int d^3\vec{y} 
[\Omega^{(2)}_\a,\Omega^{(\mathfrak{s})}_\b(\vec{y})]z^{\g_{(\mathfrak{s})}^\b}(\vec{y})
=0.
\ee
Note that
\be\label{poissonbracket-of-omega-omega}
\begin{split}
[\Omega_\a,\Omega_\g(\vec{y})]
=\lrbrk{-q_{\a\g} + \lrbrk{\frac{\pa\Omega_\a}{\pa\pa_i A_0^\g} + \frac{\pa\Omega_\g}{\pa\pa_i A_0^\a}}\pa_i}\d^{(3)}(\vec{x}-\vec{y}).
\end{split}
\ee
Imposing diffeomorphism conditions eq.\eq{diffeo-UpaiA0} and secondary-constraint enforcing relations eq.\eq{ccc},
we obtain
\be\label{poissonbracket-of-omega-omega-equal-zero}
[\Omega_\a,\Omega_\g(\vec{y})]
=0.
\ee

Next, after expressing the Poisson's brackets between primary and secondary constraints and substituting this along with eq.\eq{poissonbracket-of-omega-omega-equal-zero} into eq.\eq{1st-result-eqFJ}, one obtains

\be\label{eqC}
\cC_{0\a\g}z^{\tilde{\g}^\g}+\cC_{1\a\g}^{i}\pa_iz^{\tilde{\g}^\g}+\cC_{2\a\g}^{ij}\pa_i\pa_jz^{\tilde{\g}^\g}=0,
\ee
where
\be\label{c0}
\begin{split}
	\cC_{0\a\g}&\equiv\frac{\pa\tilde{\Omega}_\g}{\pa A_0^\a}
	-\pa_i\lrbrk{\frac{\pa\tilde{\Omega}_\g}{\pa\pa_i A_0^\a}}
	+\pa_i\pa_j\lrbrk{\frac{\pa\tilde{\Omega}_\g}{\pa\pa_i\pa_jA_0^\a}}
	\\
	&\quad
	-\lrbrk{\frac{\pa\Omega_\a}{\pa A_k^\b}
		+\frac{\pa\Omega_\a}{\pa\pa_i A_k^\b}\pa_i}\lrbrk{\frac{\pa\tilde{\Omega}_\g}{\pa\p_\b^k}
		-\pa_j\lrbrk{\frac{\pa\tilde{\Omega}_\g}{\pa\pa_j\p_\b^k}}},
\end{split}
\ee
\be\label{c1i}
\begin{split}
	\cC_{1\a\g}^{i}&\equiv 
	-\frac{\pa\tilde{\Omega}_\g}{\pa\pa_iA_0^\a}
	+2\pa_j\lrbrk{\frac{\pa\tilde{\Omega}_\g}{\pa\pa_i\pa_jA_0^\a}}
	+\frac{\pa\Omega_\a}{\pa A_k^\b}\frac{\pa\tilde{\Omega}_\g}{\pa\pa_i\p_\b^k}
	\\
	&\qquad
	-\frac{\pa\Omega_\a}{\pa\pa_iA_k^\b}\lrbrk{\frac{\pa\tilde{\Omega}_\g}{\pa\p_\b^k}
		-\pa_j\lrbrk{\frac{\pa\tilde{\Omega}_\g}{\pa\pa_j\p_\b^k}}}
	+\frac{\pa\Omega_\a}{\pa\pa_jA_k^\b}\pa_j\lrbrk{\frac{\pa\tilde{\Omega}_\g}{\pa\pa_i\p_\b^k}},
\end{split}
\ee
\be\label{c2ij}
\cC_{2\a\g}^{ij}\equiv\frac{\pa\tilde{\Omega}_\g}{\pa\pa_i\pa_jA_0^\a}
+\frac{\pa\Omega_\a}{\pa\pa_{(i|}A_k^\b}\frac{\pa\tilde{\Omega}_\g}{\pa\pa_{|j)}\p_\b^k}.
\ee

It would be helpful to rewrite eq.\eq{c0}-\eq{c2ij} in the forms which are easier to use. In particular, one may express $\cC_{0\a\g}, \cC_{1\a\g}^i, \cC_{2\a\g}^{ij}$ in terms of $\cT$ and $U_\b.$ However, even with the help of diffeomorphism invariance requirements, the expressions are still not simple to use. It is in fact even better to express these quantities in tangent bundle. We will postpone the presentation of these forms to section \ref{sec:checks}, where the relevant expressions are given in eq.\eq{C0-simple}-\eq{C1-simple}. Nevertheless, we may readily note here that by working in phase space and using diffeomorphism invariance requirements, it can be seen explicitly that
\be\label{C1-C2-properties}
\cC^i_{1\a\g} = -\cC^i_{1\g\a},\qquad
\cC^{ij}_{2\a\g} = 0.
\ee

After using eq.\eq{C1-C2-properties}, it can be seen that 
eq.\eq{eqC} becomes 
\be\label{C0C1}
\cC_{0\a\g}z^{\tilde{\g}^\g}+\cC_{1\a\g}^{i}\pa_iz^{\tilde{\g}^\g}=0.
\ee
It is clear that $z^{\tilde{\g}^\g} = 0$ is a solution to eq.\eq{C0C1}. However, the question is whether this solution is unique.
If $z^{\tilde{\g}^\g} = 0$ is the unique solution to eq.\eq{C0C1}, then after substituting into eq.\eq{2nd-result-eqFJ}, we obtain
\be\label{C0C1-alt}
(\cC_{0\g\a}-\pa_i\cC^i_{1\g\a})z^{\g^\g}
+\cC^i_{1\a\g}\pa_i z^{\g^\g} = 0.
\ee
As to be discussed in section \ref{sec:checks},
it can be shown by using diffeomorphism conditions that
\be\label{C0C1-reln}
\cC_{0\a\g} - \cC_{0\g\a} = \pa_i\cC^i_{1\a\g}.
\ee
So eq.\eq{C0C1-alt} is equivalent to eq.\eq{C0C1}.
If eq.\eq{C0C1} has the unique solution $z^{\tilde{\g}^\g} = 0,$
then $z^{\g^\g} = 0$ should also be the unique solution to eq.\eq{C0C1-alt}.
Then by using eq.\eq{zpi-eq}-\eq{za-eq}
we obtain $z^{A_\m^\a} = z^{\p^\m_\a} = 0.$
So there is no zero mode, and the procedure terminates.
By using the criteria presented by \cite{Rodrigues:2018ioe}, it can be concluded that the number of degrees of freedom is $3n$ as required.

For definiteness, let us call the condition
\be\label{C0C1-cond}
\cC_{0\a\g}z^{\tilde{\g}^\g}+\cC_{1\a\g}^{i}\pa_iz^{\tilde{\g}^\g}=0\implies \textrm{unique solution }z^{\tilde{\g}^\g} = 0
\ee
as the ``completion requirement''
since it signals the end of the second iteration. 
There are two main cases which would satisfy the completion requirement \eq{C0C1-cond}:
\bit
\item Case 1: $\cC_{1\a\g}^{i} \neq 0,$ and the boundary condition that fields should vanish fast enough near spatial infinity (this is the boundary condition which is required in the whole analysis to make integrals of total derivatives vanish) is sufficient to fix the solution to the equation in \eq{C0C1-cond} to be unique.
\item Case 2: $\cC_{1\a\g}^{i} = 0$ and $\det(\cC_{0\a\g})\neq 0.$
\eit
In the case where $\cC_{1\a\g}^{i} \neq 0,$ it is not clear whether the boundary condition would be sufficient to fix the solution to the equation in \eq{C0C1-cond} to be unique. We expect that the analysis should be done separately for each given specific theory. Even then, it would still be quite difficult, if at all possible, to show that the solution is unique.
This means that it would not be simple to show whether a given theory with $\cC_{1\a\g}^{i} \neq 0$ is within the case 1.
As for the case where a theory has $\cC_{1\a\g}^{i} = 0,$ it could be very likely that $\det(\cC_{0\a\g})\neq 0.$ This is because the form of $\cC_{0\a\g}$ contains many terms in the expression, which make it difficult for $\cC_{0\a\g}$ to be singular. On the other hand,
the requirement $\cC_{1\a\g}^{i} = 0$ itself would look quite restrictive, which might bring an immediate question as to whether it is possible to find theories within case 2. In fact, as to be explicitly discussed in subsection \ref{subsec:samples}, theories passing this requirement have already appeared in the literature. However, some of them might have been mistakenly ruled out due to the usage of the incorrect version of secondary-constraint enforcing relations \cite{ErrastiDiez:2019trb}, \cite{ErrastiDiez:2019ttn}. We will only provide one such example.

\subsubsection{Matrix form of $\cF$}
In Faddeev-Jackiw constraint analysis, it is often convenient to consider the matrix form of $\cF$. This would allow us to cross-check the analysis at the second iteration and at the same time further justify the completion requirement \eq{C0C1-cond}.
In order to obtain the components of $\cF,$ it is convenient to first denote
\be\label{yeqvf}
f_{\xi^I}
\equiv
i_{\frac{\d}{\d\xi^I}}\cF.
\ee
From direct calculation, we obtain
\be\label{yA}
f_{A_\m^\a}
=
-\d\p^\m_\a +
\sum_{\mathfrak{s} = 1}^2\int d^3\vec{y}
\frac{\d \Omega^{(\mathfrak{s})}_\b(\vec{y})}{\d A_\m^\a}\d\g_{(\mathfrak{s})}^\b(\vec{y}),
\ee
\be\label{ypi}
f_{\p^\m_\a}
=
\d A_\m^\a +
\sum_{\mathfrak{s} = 1}^2\int d^3\vec{y}
\frac{\d \Omega^{(\mathfrak{s})}_\b(\vec{y})}{\d\p^\m_\a}\d\g_{(\mathfrak{s})}^\b(\vec{y}),
\ee
\be\label{yg}
f_{\g^\a}
=-\int d^3\vec{y}\frac{\d\Omega_\a}{\d A_\m^\b(\vec{y})}\d A_\m^\b(\vec{y})
-\int d^3\vec{y}\frac{\d\Omega_\a}{\d \p^\m_\b(\vec{y})}\d \p^\m_\b(\vec{y}),
\ee
\be\label{ygt}
f_{\tilde\g^\a}
=-\int d^3\vec{y}\frac{\d\tilde{\Omega}_\a}{\d A_\m^\b(\vec{y})}\d A_\m^\b(\vec{y})
-\int d^3\vec{y}\frac{\d\tilde{\Omega}_\a}{\d \p^\m_\b(\vec{y})}\d \p^\m_\b(\vec{y}).
\ee
The matrix element of $\cF$
can then be obtained by taking interior product of eq.\eq{yA}-\eq{ygt}
with respect to phase space coordinate basis
as follows
\be
\cF_{IJ}(\vec{x},\vec{x}')=
i_{\frac{\d}{\d\xi^J(\vec{x}')}}f_{\xi^I}(\vec{x}).
\ee

The matrix form of $\cF$ is given by
\be\label{block-cF}
\cF(\vec{x},\vec{x}')
=\begin{pmatrix}
	A(\vec{x},\vec{x}') & B(\vec{x},\vec{x}')\\
	C(\vec{x},\vec{x}') & D(\vec{x},\vec{x}')
\end{pmatrix},
\ee
where
\be
A(\vec{x},\vec{x}')
=
\begin{pmatrix}
	0 & -\d_\a^\b\d_\n^\m\\
	\d_\b^\a\d_\m^\n & 0
\end{pmatrix}
\d^{(3)}(\vec{x}-\vec{x}'),
\ee
\be
B(\vec{x},\vec{x}')
=
\begin{pmatrix}
	\frac{\pa\Omega_\b}{\pa\pa_\cI A_\m^\a}(\vec{x}')&
	\frac{\pa\tilde\Omega_\b}{\pa\pa_\cI A_\m^\a}(\vec{x}')
	\\
	\frac{\pa\Omega_\b}{\pa\pa_\cI \p^\m_\a}(\vec{x}')&
	\frac{\pa\tilde\Omega_\b}{\pa\pa_\cI \p^\m_\a}(\vec{x}')
\end{pmatrix}
\pa'_{\cI}\d^{(3)}(\vec{x}-\vec{x}'),
\ee
\be
C(\vec{x},\vec{x}')
=-
\begin{pmatrix}
	\frac{\pa\Omega_\a}{\pa\pa_\cI A_\n^\b}&
	\frac{\pa\Omega_\a}{\pa\pa_\cI \p^\n_\b}\\
	\frac{\pa\tilde\Omega_\a}{\pa\pa_\cI A_\n^\b}&
	\frac{\pa\tilde\Omega_\a}{\pa\pa_\cI \p^\n_\b}
\end{pmatrix}
\pa_{\cI}\d^{(3)}(\vec{x}-\vec{x}'),
\ee
\be
D(\vec{x},\vec{x}')
=
\begin{pmatrix}
	0 & 0\\
	0 & 0
\end{pmatrix}
,
\ee
where $\pa'_{\cI}$ are generalised derivatives with respect to $\vec{x}'.$
One important steps of Faddeev-Jackiw constraint analysis is to find the determinant of $\cF.$ This determinant would also be useful when working out path integral quantisation as its square root would appear in the path integration measure.
By the standard formula of determinant of block matrix, we have
\be
\det\cF = \det(A)\det(D - CA^{-1}B).
\ee
By direct calculation, it can be shown that $\det(A) = 1.$
So in order to evaluate $\det\cF,$ one needs to first compute
$(D-CA^{-1}B).$
Direct computation gives, after applying eq.\eq{C1-C2-properties} and eq.\eq{C0C1-reln},
\be
\begin{split}	
	(D-CA^{-1}B)(\vec{x},\vec{x}')&=
	\begin{pmatrix}
		[\Omega_\a,\Omega_\b(\vec{x}')] & [\Omega_\a,\tilde\Omega_\b(\vec{x}')]\\
		[\tilde\Omega_\a,\Omega_\b(\vec{x}')] & [\tilde\Omega_\a,\tilde\Omega_\b(\vec{x}')]
	\end{pmatrix}
	\\
	&\quad
	=
	\begin{pmatrix}
		0 & -\cC_{0\a\b}-\cC^{i}_{1\a\b}\pa_i\\
		\cC_{0\b\a}-\cC^{i}_{1\b\a}(\vec{x}')\pa_i & \cD^{\cI}_{\a\b}\pa_{\cI}
	\end{pmatrix}
	\d^{(3)}(\vec{x}-\vec{x}'),
\end{split}
\ee
where $\cD^\cI_{\a\b}$ are functions whose form are not relevant to the analysis of this paper, so we do not provide its explicit form.

In order for $(D-CA^{-1}B)$ to be invertible,
the solution $w$ of $(D-CA^{-1}B)w = \psi$ should be unique.
Let us denote $w(\vec{x}')
\equiv (u^{\b}(\vec{x}'), v^{\b}(\vec{x}'))^T,$
and $\psi(\vec{x})\equiv (\chi_{\a}(\vec{x}), \l_{\a}(\vec{x}))^T.$
So
\be\label{eq-v}
-\lrbrk{\cC_{0\a\b} + \cC^i_{1\a\b}\pa_i}v^\b = \chi_\a,
\ee
\be\label{eq-u}
(\cC_{0\a\b} + \cC^i_{1\a\b}\pa_i)u^\b+
\cD^{\cI}_{\a\b}\pa_{\cI}v^\b
=\l_\a.
\ee
Solution to eq.\eq{eq-v} is
\be
v^\b
=\int d^3\vec{x}'G^{\b\g}(\vec{x},\vec{x}')\chi_\g(\vec{x}') + v_0^\b,
\ee
where $G^{\b\g}(\vec{x},\vec{x}')$ and $v_0^\b$ satisfy
\be
-\lrbrk{\cC_{0\a\b} + \cC^i_{1\a\b}\pa_i}G^{\b\g}(\vec{x},\vec{x}')
=\d_\a^\g\d^{(3)}(\vec{x}-\vec{x}'),
\ee
and
\be
-\lrbrk{\cC_{0\a\b} + \cC^i_{1\a\b}\pa_i}v_0^\b = 0.
\ee
In order for $v^\b$ to be the unique solution to eq.\eq{eq-v}, we demand that $v_0^\b$ is unique. This is precisely the completion requirement
\eq{C0C1-cond}.

In the case where $D-CA^{-1}B$ is invertible, the determinant of $\cF$ can be determined. In this case, by direct calculation using the standard formula of determinant of block matrix and using the property of determinant of product of square matrices, one obtains
\be\label{detF}
\det\cF = \{\det[(\cC_{0\a\b}+\cC_{1\a\b}^i\pa_i)\d^{(3)}(\vec{x}-\vec{x}')]\}^2.
\ee
Demanding that there is no zero mode of $\cF$ at the second iteration is equivalent to demanding that $\det\cF\neq 0.$ So by using eq.\eq{detF}, it can be seen that one should demand the differential operator $\cC_{0\a\b}+\cC_{1\a\b}^i\pa_i$ to have no zero mode. This also implies the completion requirement.

The class of theories we consider indeed include the particular theories investigated in \cite{ErrastiDiez:2020dux}, in which the conditions called ``quantum consistency condition'' are derived.
Based on the result of our paper, these conditions can indeed be generalised to a larger class of theories.
The generalisation is simply the condition we called ``completion requirement''.
The idea is that our differential operator $\cC_{0\a\b}+\cC_{1\a\b}^i\pa_i$ could be thought of as a generalisation to their differential operator $Z_{\a\b}.$ 
We have provided in eq.\eq{C0-simple}-\eq{C1-simple} the formula to directly compute the coefficients $\cC_{0\a\b}$ and $\cC_{1\a\b}^i,$
which in turn give rise the required differential operator.
The quantum consistency condition derived in
\cite{ErrastiDiez:2020dux} is $Z_{\a\b}\neq 0.$ This seems to demand a differential operator to be non-zero.
We suppose that it would be useful to give a slightly clearer interpretation.
In particular, one should interpret it as being that the differential operator $Z_{\a\b}$ has no zero mode. This is exactly generalised to our requirement.

Furthermore, by using diffeomorphism invariance requirement, we have shown that $\cC_{0\b\a} = \cC_{0\a\b} - \pa_i\cC^i_{1\a\b}$ and $\cC^i_{1\a\b} = -\cC^i_{1\b\a}.$ This implies that $\cC_{0\b\a}-\cC_{1\b\a}^i(\vec{x}')\pa_i = \cC_{0\a\b}+\cC_{1\a\b}^i\pa_i,$ which should be the generalisation to $-Z'_{\b\a} = Z_{\a\b}$ of the theories in \cite{ErrastiDiez:2020dux}.
This provides an explanation why the determinant of the symplectic two-form factorises as eq.\eq{detF}. For example, in the particular theories of \cite{ErrastiDiez:2020dux}, the determinant reduces as $\det\cF = (\det(Z\d^{(3)}(\vec{x}-\vec{x}')))^2 = \det(Z\cdot Z\d^{(3)}(\vec{x}-\vec{x}'))
= \det(-Z'\cdot Z\d^{(3)}(\vec{x}-\vec{x}')),$ in agreement, modulo a possible minor typographical error, with 
\cite{ErrastiDiez:2020dux}.

An immediate application is that
if the theory passes the completion requirement,
path integral quantisation can be carried out
\cite{Toms:2015lza}.
In particular,
it is possible to read off
\be
\sqrt{\det\cF} = \det[(\cC_{0\a\b}+\cC_{1\a\b}^i\pa_i)\d^{(3)}(\vec{x}-\vec{x}')],
\ee
which is an expression that
appears in the measure of the generating functional in path integral quantisation.

\section{\label{sec:checks}Consistency check using Lagrangian constraint analysis}
In the previous section, we have presented the criteria
for which the theories of $n$ vector fields with Lagrangian of the form eq.\eq{L-UV} would have $3n$ degrees of freedom,
which corresponds to theories of multi-field generalised Proca.
In short, the criteria is that the theory should transform in a standard way under diffeomorphism transformation; it should satisfy eq.\eq{HssnconT}, \eq{ccc}, and \eq{C0C1-cond}.

In this section, we present a consistency check of our result by using Lagrangian constraint analysis developed in \cite{ErrastiDiez:2019trb}, \cite{ErrastiDiez:2019ttn}, \cite{ErrastiDiez:2020dux}, \cite{Kamimura:1981fe}, and work out the equivalence between the conditions to be obtained in this section with those from the previous section.

In this analysis, it is convenient to define collective coordinates as follows.
Let $Q^M, Q^\a, Q^A$ be collective for $A_\m^\a, A_0^\a, A_i^\a,$ respectively. The Lagrangian we are intereseted in is given by
\be
\cL = \cL(Q^M,\dot Q^M, \pa_i Q^M, K),
\ee
Euler-Lagrange equations for vector fields are
\be\label{EL-Q}
\begin{split}
	0
	&=\frac{d}{dt}\lrbrk{\frac{\pa\cL}{\pa\dot Q^M}}
	+\pa_i\lrbrk{\frac{\pa\cL}{\pa\pa_i Q^M}}
	-\frac{\pa\cL}{\pa Q^M}\\
	&=W_{MN}\ddot{Q}^N + \a_M,
\end{split}
\ee
where
\be\label{WMN}
W_{MN}\equiv
\frac{\pa^2\cL}{\pa\dot Q^M\pa\dot Q^N},
\ee
\be\label{aM}
\begin{split}
	\a_M
	&=\frac{\pa^2\cL}{\pa\dot Q^M\pa Q^N}\dot Q^N
	+\frac{\pa^2\cL}{\pa\dot Q^M\pa\pa_i Q^N}\pa_i\dot Q^N+\frac{\pa^2\cL}{\pa\dot Q^M\pa K}{\dot K}
	+\pa_i\lrbrk{\frac{\pa\cL}{\pa\pa_i Q^M}}
	\\
	&\qquad
	-\frac{\pa\cL}{\pa Q^M}.
\end{split}
\ee
The special Hessian conditions eq.\eq{Hssncon} give the following conditions on $W_{MN}:$
\be
W_{\a N} = 0,\qquad
\det(W_{AB}) \neq 0.
\ee
So Euler Lagrange equations \eq{EL-Q} can be separated into
equations of motion:
\be
W_{AB}\ddot Q^B + \a_A = 0,
\ee
and primary constraints
\be
\a_\a = 0.
\ee
Let $M^{AB}$ be the inverse of $W_{AB}.$ 
So the equations of motion imply
\be\label{eomMAB}
\ddot Q^A + M^{AB}\a_B = 0.
\ee

Time evolution of constraints is given by, after making use of eq.\eq{eomMAB},
\be\label{dotaa}
\begin{split}
	\dot{\a}_\a
	&=
	\sum_{|\cI|=0}^1\frac{\pa\a_\a}{\pa\pa_\cI\dot Q^\b}\pa_\cI\ddot Q^\b
	+\sum_{|\cI|=0}^2\frac{\pa\a_\a}{\pa\pa_\cI Q^M}\pa_\cI\dot Q^M
	-\sum_{|\cI|=0}^1\frac{\pa\a_\a}{\pa\pa_\cI\dot Q^B}\pa_\cI(M^{BC}\a_C)
		\\
	&\qquad
	+\frac{\pa\a_\a}{\pa K}\dot K.
\end{split}
\ee
We demand that the process should not terminate at this stage. So
the conditions $\dot{\a}_\a = 0$ should not introduce further dynamics on the vector fields. This means that the expressions with second order derivative in time of $Q^\b$ should not appear in eq.\eq{dotaa}. These expressions are $\ddot Q^\b$ and $\pa_i\ddot Q^\b.$
From direct calculation, their coefficients are
\be\label{ccc-alt}
\frac{\pa\a_\a}{\pa\dot Q^\b}
\equiv
\frac{\pa^2\cL}{\pa\dot Q^\a\pa Q^\b}
-\frac{\pa^2\cL}{\pa Q^\a\pa\dot Q^\b}
+\pa_i\lrbrk{\frac{\pa^2\cL}{\pa\pa_i Q^\a\pa\dot Q^\b}},
\ee
and
\be
\frac{\pa\a_\a}{\pa\pa_i\dot Q^\b}
=\frac{\pa^2\cL}{\pa\dot Q^\a\pa\pa_i Q^\b}
+\frac{\pa^2\cL}{\pa\dot Q^\b\pa\pa_i Q^\a}.
\ee
By using a diffeomorphism condition \eq{diffeo-UpaiA0},
the coefficient of $\pa_i\ddot Q^\b$ vanishes. So we are left with the terms with $\ddot Q^\b$. In order for the coefficients of these terms to vanish, we should set
\be\label{paapadotQ}
\frac{\pa\a_\a}{\pa\dot Q^\b}
=0,
\ee
which turns out to be equivalent to eq.\eq{ccc}.

Two remarks are in order. The first is that the analysis in \cite{ErrastiDiez:2019trb} does not show explicit dependence on spatial derivatives of fields. While this might be sufficient for the purpose of counting the number of degrees of freedom, the conditions derived in the process are not readily correct until time dependence on spatial derivatives of fields are re-introduced. From their analysis, the last term on RHS of eq.\eq{ccc-alt} is missing. This term could be considered as restoring spatial derivatives of fields. The second remark is that the reference \cite{ErrastiDiez:2020dux} does not seem to mention the dependence of $\dot\a_\a$ on $\pa_i\ddot Q^\b$ nor on
whether their coefficients disappear. We have learned from the analysis above that diffeomorphism invariance requirement is crucial, at least in the case of multi-field generalised Proca theories that we are analysing, to make the the coefficients disappear. It would be interesting to see whether this behaviour is also the case in the analysis of more general theories given in \cite{ErrastiDiez:2020dux}.

Although Lagrangian constraint analysis is more advantageous than Hamiltonian constraint analysis in that it treats time and space on a more equal footing, the nature of constraint analysis still requires that time and space should be treated differently. For example, to see whether there are further constraints, only the time evolution is required. Some information on manifest covariance 
would then be lost. In order to recover them, one needs to make use of the fact that theories are diffeomorphism invariance (or, in the case of flat spacetime, Lorentz invariance).

Let us continue the analysis. By imposing eq.\eq{paapadotQ},
we then have $n$ secondary constraints $\f_\a = \dot\a_\a \approx 0.$ The next step is to consider the time evolution of $\f_\a.$ We demand that the condition $\dot{\f}_\a\approx 0$ should not lead to further constraints. For this, $\dot{\f}_\a$ should contain terms with second order derivative in time on $Q^\b.$ These terms are
\be
\frac{\pa\f_\a}{\pa\dot Q^\b}\ddot Q^\b
+\frac{\pa\f_\a}{\pa\pa_i\dot Q^\b}\pa_i\ddot Q^\b
+\frac{\pa\f_\a}{\pa\pa_i\pa_j\dot Q^\b}\pa_i\pa_j\ddot Q^\b
\in
\dot\f_\a.
\ee
The analysis in \cite{ErrastiDiez:2020dux} does not mention terms with $\pa_i\ddot Q^\b$ and $\pa_i\pa_j\ddot Q^\b.$ In principle, these terms are also crucial in determining whether the procedure should be terminated. Analysis of a particular case, for example in \cite{ErrastiDiez:2021ykk}, also show the dependence of constraints on these terms, especially $\pa_i\ddot Q^\b.$

Let us connect the result in this subsection with the analysis in phase space given in section \ref{sec:analysis}. For this, we first show that by transforming to tangent bundle, $\tilde{\Omega}_\a = -\a_\a.$
We start from eq.\eq{tildeOmega}.
Then by using $T=\cL-U_\g\dot{Q}^\g$, and realising that $U_\a$ is independent of $\dot{Q}^M$, we obtain\footnote{It is understood that LHS of eq.\eq{tildeOmega-alpha} is actually the pullback of $\tilde{\Omega}_\a$ to tangent bundle. Throughout this paper, we do not use different notations to distinguish the functions from their pullbacks as it should be clear from the context.}
\be\label{tildeOmega-alpha}
\begin{split}
	\tilde{\Omega}_\a
	&=-\a_\a
	+\lrbrk{\frac{\pa U_\a}{\pa Q^\b}-\frac{\pa U_\b}{\pa Q^\a}+\pa_i\lrbrk{\frac{\pa U_\b}{\pa\pa_iQ^\a}}}\dot{Q}^\b
	+\lrbrk{\frac{\pa U_\a}{\pa\pa_iQ^\b}+\frac{\pa U_\b}{\pa\pa_iQ^\a}}\pa_i\dot{Q}^\b.
\end{split}
\ee
The second and the third term on RHS vanish due to secondary-constraint enforcing relations \eq{ccc}
and diffeomorphism invariance requirement \eq{diffeo-UpaiA0}.
This finally gives
\be\label{Oaaa}
\tilde{\Omega}_\a=-\a_\a,
\ee
as required. Then by following the calculations outlined in Appendix \ref{app:paphipaQdot}, we obtain
\be\label{f0-f1-f2-fin}
\frac{\pa\f_\a}{\pa\dot Q^\b}
= -\cC_{0\a\b},\quad
\frac{\pa\f_\a}{\pa\pa_i\dot Q^\b}
=-\cC^{i}_{1\a\b},\quad
\frac{\pa\f_\a}{\pa\pa_i\pa_j\dot Q^\b}
=0.
\ee
Note in passing that the condition
\be
\cC_{0\a\b}
=\cC_{0\b\a}-\pa_i\cC_{1\b\a}^i,
\ee
which is also proven in Appendix \ref{app:paphipaQdot} is crucial in the derivation of eq.\eq{f0-f1-f2-fin}.

Therefore, time evolution of $\dot{\f}_\a$ is of the form
\be
\dot\f_\a
=-\lrbrk{\cC_{0\a\b}
	+\cC_{1\a\b}^i\pa_i}\ddot Q^\b
+\cdots,
\ee
where $\cdots$ are terms with up to first order in time derivative in $Q^M.$
In order for $\dot{\f}\approx 0$ not to lead to further constraints, we should demand that it is equivalent to
\be
\ddot Q^\b +\cdots = 0.
\ee
This would be possible only when 
the differential operator
\be
\lrbrk{\cC_{0\a\b}
	+\cC_{1\a\b}^i\pa_i}
\ee
is invertible. Equivalently, this differential operator should have no zero mode. This would lead exactly to the completion requirements \eq{C0C1-cond} given at the end of subsection \ref{sec:second-it}.

We have seen that the analysis of Lagrangian constraint analysis agree with the Faddeev-Jackiw constraint analysis. In particular, the functions $\cC_{0\a\b}$ and $\cC^i_{1\a\b}$ appear in ones of the important conditions. Having worked with Lagrangian analysis, we are now in a position to express them in a more useful form. They are
\be\label{C0-simple}
\cC_{0\a\b}
=-\frac{\pa\a_\a}{\pa A_0^\b}-\pa_j\lrbrk{\frac{\pa\a_\a^j}{\pa\dot A_0^\b}}
+\frac{\pa\a_\g^k}{\pa\dot A_0^\a}M^{\g\d}_{kl}\frac{\pa\a_\d^l}{\pa\dot A_0^\b},
\ee
\be\label{C1-simple}
\cC_{1\a\b}^i
=-\frac{\pa\a_\a}{\pa\pa_i A_0^\b}-\frac{\pa\a_\a^i}{\pa\dot A_0^\b} - \frac{\pa\a_\a}{\pa\dot A_i^\b},
\ee
where
\be\label{C0C1-expr1}
\frac{\pa\a_\a}{\pa A_0^\b} =
\pa_\m\lrbrk{\frac{\pa^2\cL}{\pa\pa_\m A_0^\a \pa A_0^\b}}
-\frac{\pa^2\cL}{\pa A_0^\a\pa A_0^\b},
\ee
\be\label{C0C1-expr2}
\frac{\pa\a_\g^k}{\pa\dot A_0^\b}
=\frac{\pa^2\cL}{\pa A_0^\b\pa\dot A^\g_k}
+\pa_i\lrbrk{\frac{\pa^2\cL}{\pa\dot{A}_0^\b\pa\pa_i A^\g_k}}
-\frac{\pa^2\cL}{\pa\dot{A}_0^\b\pa A^\g_k},
\ee
\be\label{C0C1-expr3}
\begin{split}
	\frac{\pa\a_\a}{\pa\pa_i A_0^\b}
	&
	=
	\pa_\m\lrbrk{\frac{\pa^2\cL}{\pa\pa_\m A_0^\a\pa\pa_i A_0^\b}}
	+2\frac{\pa^2\cL}{\pa\pa_i A_0^{[\a}\pa A_0^{\b]}}.
\end{split}
\ee

\section{\label{sec:application}Application of the sufficient conditions}

In the previous sections, we have studied a class of theories of $n$ vector fields, with a possiblity to couple to external fields. A theory in this class describes
$n-$field generalised Proca system coupled to external fields
if it passes the special Hessian condition \eq{Hssncon}, secondary-constraint enforcing relation \eq{ccc-tb}, as well as the completion requirement which demands that eq.\eq{C0C1} contains no zero mode. The completion requirement is the most involved. In order to consider them, one needs to write down the expression of $\cC_{0\a\b}$ and $\cC^i_{1\a\b}.$ Their explicit forms can be computed by using eq.\eq{C0-simple}-\eq{C0C1-expr3}.

In this section, we will demonstrate the use of the criteria presented in sections \ref{sec:analysis}-\ref{sec:checks}.
We provide a few examples of theories which pass these requirements, as well as an example theory which does not pass, but is previously incorrectly identified in the literature as being legitimate. These examples should be sufficient to serve the purpose. They are, however, far from exhaustive. We expect that many other theories passing these requirements are already presented in the literature, but some of them may have been previously misinterpreted.
\subsection{\label{subsec:samples}Examples}
\subsubsection{Separable multi-field generalised Proca theories}
One of simple examples is the case where each of the 
$n$ vector fields in the system does not couple to one another.
The system is considered to be separated into $n$ sub-systems
of single vector field, possibly coupled to external fields.
It could then be expected that one can simply separately apply the
constraint analysis on on each sub-system. For example, an analysis of 
\cite{Sanongkhun:2019ntn} confirms that as long as each sub-system describes a generalised Proca field, possibly coupled to external fields,
then the vector sector has $3$ degrees of freedom.

Direct use of the results presented in sections \ref{sec:analysis}-\ref{sec:checks} can also easily be done. The Lagrangian of the example system takes the form
\be
\cL= \sum_{\a=1}^{n}\cL_{(\a)},
\ee
where for each $\a\in\{1,2,\cdots,n\},$ the sub-Lagrangian $\cL_{(\a)}$ is a function of only the $\a$th vector field $A_\m^\a$, its first order derivative $\pa_\m A_\n^\a$, and possibly external fields; but $\cL_{(\a)}$ does not depend on the $\b$th vector fields nor their derivatives if $\b\neq \a.$
After demanding that it satisfies the special Hessian condition \eq{Hssncon}, we obtain
\be
\cL_{(\a)}
= T_{(\a)}
+U_\a\dot{A}_0^\a\quad(\text{no summation over $\a$}).
\ee
So we have 
\be
\frac{\pa U_\a}{\pa A_0^\b}=\d_{\a\b}\frac{\pa U_\a}{\pa A_0^\a} \quad(\text{no summation over $\a$}),
\ee
and from eq.\eq{diffeo-UpaiA0}, we have
\be
\frac{\pa U_\a}{\pa\pa_i A_0^\b}=0.
\ee
Therefore, the secondary-constraint enforcing relations are automatically satisfied.
Next,
since the derivative of $\cL_{(\a)}$ with respect to $A_\m^\b$ or $\pa_\m A_\n^\b$ vanish if $\a\neq \b,$ then $\cC_{0\a\b}$ and $\cC^i_{1\a\b}$ are diagonal matrices.
In fact, since $\cC^i_{1\a\b} = -\cC^i_{1\b\a},$ we can conclude that $\cC^i_{1\a\b} = 0.$
So we have
\be
\cC_{0\a\b} = \cC_{0\a\a}\d_{\a\b},\quad
\cC^i_{1\a\b} = 0,\quad \textrm{(no sum over $\a$)}.
\ee
Then in order for eq.\eq{C0C1}
to have no zero mode, we should require
\be
\det(\cC_{0\a\b})
=\prod_{\a=1}^n\cC_{0\a\a}
\neq 0,
\ee
which is possible if $\cC_{0\a\a}\neq 0$ for each $\a\in\{1,2,\cdots,n\}.$
This means that each sub-system has to be described by a generalised Proca field, possibly coupled to external fields.

\subsubsection{A less trivial example}
Let us consider an example theory whose Lagrangian is of the form
\be\label{cL2}
\cL = \cL_2(A^\a_\m,A^\a{}_{\m\n},K),
\ee
where $A^\a{}_{\m\n}\equiv \pa_\m A^\a_\n - \pa_\n A^\a_\m.$
It is one of the simplest forms of multi-field generalised Proca theories being presented in the literature, see for example \cite{ErrastiDiez:2019trb}, \cite{ErrastiDiez:2019ttn}, \cite{GallegoCadavid:2020dho}, \cite{Allys:2016kbq}.
We confirm that the theory is indeed legitimate. For this theory,
\be
\frac{\pa\cL}{\pa\pa_\m A_\n^\a}
=-\frac{\pa\cL}{\pa\pa_\n A_\m^\a}
=2\frac{\pa\cL_2}{\pa A^\a{}_{\m\n}}.
\ee
This immediately gives $U_\a = 0.$
So the secondary-constraint enforcing relations \eq{ccc-tb} is trivially satisfied. Furthermore, $\cC_{0\a\b}$ and $\cC^i_{1\a\b}$ are simplified to
\be
	\cC_{0\a\b}
	=\frac{\pa^2\cL_2}{\pa A_0^\a\pa A_0^\b}
	-4\frac{\pa^2\cL_2}{\pa A^\g{}_{0j}\pa A_0^\a}M^{\g\d}_{jk}\frac{\pa^2\cL_2}{\pa A^\d{}_{0k}\pa A_0^\b},
	\quad\cC^i_{1\a\b} = 0.
\ee
It can be seen that, apart from some exceptions, $\det\cC_{0\a\b}\neq 0.$ So the theory has the required number of degrees of freedom, and hence is an $n-$field generalised Proca theory.

A notable exception is when $\cL$ is independent from $A_0^{\a_1}$ for $\a_1\in\{1,2,\cdots,r\},$ where $1<r\leq n.$ While the criteria provided in sections \ref{sec:analysis}-\ref{sec:checks} can only be used to state that this exception is not an $n-$field generalised Proca theory, it should nevertheless intuitively be expected that it describes $(n-r)$ generalised Proca fields while the other $r$ fields might be, provided that it passes some further criteria, generalised Maxwell fields. These criteria, if any, should arise when one considers multi-field generalised Maxwell-Proca theories.
While \cite{ErrastiDiez:2019trb}, \cite{ErrastiDiez:2019ttn}, \cite{ErrastiDiez:2020dux} might have already provided the criteria for identifying multi-field generalised Maxwell-Proca theories, we have found in this work that even when restricted to purely (multi-field) Proca theories, their analysis seems to require some non-trivial refinements. So we expect that the refinements to the criteria of multi-field generalised Maxwell-Proca theories are needed. We leave this for future works.

Nevertheless, suppose that we have considered a Lagrangian $\cL^{(1)}$ whose $\cC^i_{1\a\b},$ denoted $\cC^i_{1\a\b}(\cL^{(1)}),$ is zero while its $\cC_{0\a\b},$ denoted $\cC_{0\a\b}(\cL^{(1)}),$ is singular. It could still be possible to add to it another Lagrangian $\cL^{(2)}$ with $\cC^i_{1\a\b}(\cL^{(2)}) = 0$ so that the resulting Lagrangian $\cL^{(1)}+\cL^{(2)}$ might describe an $n-$field generalised Proca theory. This is because, due to eq.\eq{C1-simple}, $\cC^i_{1\a\b}$ is linear. So $\cC^i_{1\a\b}(\cL^{(1)}+\cL^{(2)}) = \cC^i_{1\a\b}(\cL^{(1)})+\cC^i_{1\a\b}(\cL^{(2)}) = 0.$
On the other hand, due to the last term on RHS of eq.\eq{C0-simple}, $\cC_{0\a\b}$ is non-linear. So
$\cC_{0\a\b}(\cL^{(1)}+\cL^{(2)}) = \cC_{0\a\b}(\cL^{(1)})+\cC_{0\a\b}(\cL^{(2)})
+\textrm{non-linear}(\cL^{(1)},\cL^{(2)}).$
Due to non-linearity of $\cC_{0\a\b}$ and of its determinant, it is likely that $\cC_{0\a\b}(\cL^{(1)}+\cL^{(2)})$ is not singular even if both $\cC_{0\a\b}(\cL^{(1)})$ and $\cC_{0\a\b}(\cL^{(2)})$ are singular. Of course, although highly likely to be the case, direct calculations are required in each case to confirm whether this is truly the case.

\subsubsection{\label{subsec:legit}A legitimate theory previously misinterpreted}
In \cite{Allys:2016kbq}, actions for multiple vector fields are constructed by using a systematic approach which demands that the special Hessian condition is satisfied. In principle, this is not sufficient to give legitimate theories as further conditions, for example secondary-constraint enforcing relations, are required. The reference \cite{ErrastiDiez:2019trb} points out that
one of theories proposed in \cite{Allys:2016kbq}, does not pass secondary-constraint enforcing relations and hence contains extra degrees of freedom. The Lagrangian of this theory is
\be\label{legit-su2}
	\cL = -\ove{4}A^\a{}_{\m\n} A_\a{}^{\m\n} 
	-4\l\bigg(A^{\a\s}A^\b_\s\pa^{\m}A^\a_{[\m}\pa^{\n}A^\b_{\n]}
	+A^\a_{[\m}A^\b_{\n]}\pa^{\m}A^\a_\r\pa^\n A^{\b\r}\bigg),
\ee
where $\l$ is a non-zero constant.
Actually, since secondary-constraint enforcing relations presented in \cite{ErrastiDiez:2019trb} miss some terms in the expression, in principle, the interpretation being drawn should be revised.

Let us argue that in fact the theory \eq{legit-su2} is legitimate.
By direct calculation, one obtains
\be
	\frac{\pa^2\cL}{\pa\dot A^\a_0\pa A^\b_0}-\frac{\pa^2\cL}{\pa\dot A^\b_0\pa A^\a_0}
	=-8\l\pa_i(A_{[0}^{\a} A_{i]}^{\b})
	=-\pa_i\lrbrk{\frac{\pa^2\cL}{\pa\dot A_0^\b\pa\pa_i A_0^\a}},
\ee
which means that the secondary-constraint enforcing relation \eq{ccc-tb} is satisfied. Therefore, contrary to the interpretation given in \cite{ErrastiDiez:2019trb}, the theory eq.\eq{legit-su2} has secondary constraints. Furthermore, this theory is in fact an $n-$field generalised Proca theory. To see this, one notes that by making direct computation one obtains
\be
\cC_{1\a\b}^i = 0.
\ee
It can then be checked that if $\l\neq 0,$
then $\det(\cC_{0\a\b})\neq 0.$ Therefore, the completion requirement \eq{C0C1-cond} is satisfied.

Of course, the same conclusion can also be reached if one directly starts from the Lagrangian \eq{legit-su2} and performs either Hamiltonian or Lagrangian constraint analysis.

We expect that there are also other theories presented in \cite{Allys:2016kbq} which are legitimate but is previously incorrectly ruled out. A common feature for these theories is that
\be\label{non-ccc-old}
\frac{\pa^2\cL}{\pa\dot A^\a_0\pa A^\b_0}-\frac{\pa^2\cL}{\pa\dot A^\b_0\pa A^\a_0}
\neq 0,
\ee
which makes them incorrectly ruled out.
So if $\pa^2\cL/(\pa\dot A_0^\b\pa\pa_i A_0^\a)\neq 0,$ then one might try to see if $-\pa_i(\pa^2\cL/(\pa\dot A_0^\b\pa\pa_i A_0^\a))$ would cancel out with LHS of \eq{non-ccc-old}. If this is the case, then one can proceed to check the completion requirement.

\subsubsection{An undesired theory previously misinterpreted}
After the reference \cite{ErrastiDiez:2019trb} suggests 
that the special Hessian conditions are not sufficient, and that the secondary-constraint enforcing relations should be satisfied, theories are being proposed in the literature in order to satisfy the required relations. Notable examples are \cite{ErrastiDiez:2019trb}, \cite{ErrastiDiez:2019ttn}, \cite{GallegoCadavid:2020dho}.

Let us argue that, by using a refined version of secondary-constraint enforcing relations, some of the theories in fact are undesired, i.e. they contain extra degrees of freedom. In particular, we explicitly show one example from \cite{GallegoCadavid:2020dho}. This particular example has the Lagrangian of the form
\be
\cL\! =\! -2A^\a{}_{\m\n} S^{\b\m}{}_\s A_{\a\r}A_{\b\l}\e^{\n\s\r\l}
+S^\a{}_{\m\n}S^{\b\n}{}_\s A_{\a\r}A_{\b\l}\e^{\m\s\r\l},
\ee
where $S^\a{}_{\m\n}\equiv\pa_\m A_\n^\a + \pa_\n A_\m^\a.$
By direct calculation, one obtains
\be
\frac{\pa^2\cL}{\pa\dot A^\a_0\pa A^\b_0}-\frac{\pa^2\cL}{\pa\dot A^\b_0\pa A^\a_0}
=0
\neq-\pa_i\lrbrk{\frac{\pa^2\cL}{\pa\dot A_0^\b\pa\pa_i A_0^\a}}.
\ee
Therefore, this theory is in fact undesired.

We expect that there are also other theories presented in the literature which contain extra degrees of freedom but is previously interpreted as being well-behaved. For these theories,
$\pa^2\cL/(\pa\dot A^\a_0\pa A^\b_0) -\pa^2\cL/(\pa\dot A^\b_0\pa A^\a_0)
=0.
$
So if they are truly undesired,
one should find that
$-\pa_i(\pa^2\cL/(\pa\dot A_0^\b\pa\pa_i A_0^\a))\neq 0,$
which would violate the secondary-constraint enforcing relations \eq{ccc-tb}.
\subsection{\label{subsec:implications}Cosmological implications}

Multi-field generalised Proca theories have been applied
for example in \cite{Rodriguez:2017wkg}, \cite{Gomez:2019tbj}, \cite{Garnica:2021fuu} to explain cosmological phenomena. In some of these studies, the conditions presented by \cite{ErrastiDiez:2019trb}, \cite{ErrastiDiez:2019ttn} are taken into consideration. However, as we have been discussing, these conditions are incorrect and should be replaced by eq.\eq{ccc-tb}. In principle, one should then investigate the validation of the cosmological implications presented in \cite{Rodriguez:2017wkg}, \cite{Gomez:2019tbj}, \cite{Garnica:2021fuu}. In this subsection, we discuss a direction for further investigations on these works.

In \cite{Rodriguez:2017wkg}, a Lagrangian involving Einstein-Hilbert term, $SU(2)$ Yang-Mills term $\cL_{YM}$, and a term called $\a\cL_4^1$ where $\a$ is a constant
is considered. Autonomous dynamical system analysis of this model in a homogeneous and isotropic background is studied which allows dark energy and primordial inflation to be discussed. While the dark energy case leads to an interesting result, the primordial inflation case is problematic as the model is strongly sensitive to initial conditions and the value of $\a.$ It is then suggest that one should also include a term $\k\cL_4^2,$ where $\k$ is a constant, into the Lagrangian and see if the problem can be evaded.

Let us discuss whether the Lagrangian presented in \cite{Rodriguez:2017wkg} would pass the sufficient conditions in section \ref{sec:analysis}. Note that for the theory in \cite{Rodriguez:2017wkg}, gravity is dynamical whereas the sufficient conditions we have presented is useful when the gravity is non-dynamical. Nevertheless, a simple check can still be performed in the case of flat spacetime, in which case $\cL_{YM}$ is a function of $A^\a_\m, A^\a{}_{\m\n},$ whereas $\cL_4^1$ is a function of $A^\a_\m, \pa_\m A^\a_{\n}$ in such a way that $\pa^2\cL_4^1/\pa\dot A_0^\a\pa A_0^\b = \pa^2\cL_4^1/\pa\dot A_0^\b\pa A_0^\a,$ $\pa^2\cL_4^1/(\pa \dot A_0^\a\pa\pa_i A_0^\b) = 0.$
So it can easily be seen from the discussion of subsection \ref{subsec:samples} that the theory in \cite{Rodriguez:2017wkg} pass the sufficient conditions.

It would also be interesting to investigate whether the suggestion to include the term $\k\cL_4^2$ still valid, as far as our sufficient conditions are concerned. So let us also consider the case of flat spacetime. In this case, it can easily be seen that $\cL_{YM}+\k\cL_4^2$ is simply expressible as a summation of the Lagrangians \eq{cL2} and \eq{legit-su2}.
So indeed the term $\k\cL_4^2$ can be included to extend the model of \cite{Rodriguez:2017wkg}. Note on the other hand that if one had used the criteria of \cite{ErrastiDiez:2019trb}, \cite{ErrastiDiez:2019ttn}, the term $\k\cL_4^2$ would have been incorrectly ruled out.

In \cite{Gomez:2019tbj}, \cite{Garnica:2021fuu}, cosmological implications of multi-field generalised Proca theories are also investigated. It turns out however that some terms of the Lagrangian, for example $\cL_4^2$ presented in \cite{Rodriguez:2017wkg}, has been incorrectly ruled out according to the criteria of \cite{ErrastiDiez:2019trb}, \cite{ErrastiDiez:2019ttn}. But as discussed in the previous paragraph, such a term in fact passes the criteria presented in section \ref{sec:analysis}, so there is no problem with the number of degrees of freedom.
It would be interesting to see for example the cosmological implication of the inclusion of $\cL_4^2$ to the models of 
\cite{Gomez:2019tbj}, \cite{Garnica:2021fuu}.

\section{\label{sec:conclusion}Discussion and Conclusion}
In this work, we have worked out the sufficient conditions to make a theory describe multi-field generalised Proca theory, possibly coupled to external fields. We focus on a class of theories whose Lagrangians are functions of up to first-order derivative of the vector fields. Furthermore, we demand that the Lagrangian of each theory satisfies the special Hessian condition \eq{Hssncon}, free of Ostrogradski instability and that it transforms in a standard way under standard diffeomorphism. Theories in this class should also pass the secondary-constraint enforcing relations \eq{ccc} (or equivalently, eq.\eq{ccc-tb}) as well as the completion requirements \eq{C0C1-cond} which can be computed using eq.\eq{C0-simple}-\eq{C0C1-expr3}.

We have obtained these conditions by using Faddeev-Jackiw constraint analysis and cross checked using Lagrangian constraint analysis. In the analysis, diffeomorphism invariance requirements, eq.\eq{diffeo-UpaiA0}, \eq{cond-A0dot-phasespace-1}-\eq{cond-A0dot-phasespace-2} are needed.
The diffeomorphism invariance requirements are not extra conditions.
They are in fact conditions for which every diffeomorphism invariance theory is satisfied.
If one analyses each specific theory one by one,
it can be explicitly seen that these requirements are automatically satisfied.
However, if one analyses a class of theories at a time, diffeomorphism invariance is less manifest as, by the nature of constraint analysis, time and space are not treated on equal footing.
In this case, diffeomorphism invariance requirements
help to realise the diffeomorphism invariance that every theory in the class  possesses.
These requirements are especially useful in simplifying key expressions in intermediate steps. Let us provide two example instances where the usefulness of diffeomorphism invariance requirements when analysing a class of theories are shown.

The first example is that, if the secondary-constraint enforcing relations \eq{ccc} is imposed, and if one does not know that theories which are diffeomoprhism invariant should satisfy eq.\eq{diffeo-UpaiA0}, one would not be able to see, when analysis a class of theories, that eq.\eq{pre-ccc} is trivial, and hence would impose eq.\eq{diffeo-UpaiA0} as another, but is in fact obsolete, 
secondary-constraint enforcing relations.
Another notable example is that diffeomorphism invariance requirements allow us to realise the connection between results from Faddeev-Jackiw constraint analysis and Lagrangian constraint analysis. The diffeomorphism invariance requirements have been helping in simplifying $\cC_{0\a\b}, \cC^i_{1\a\b}, \cC^{ij}_{2\a\b}$ and allowing us to realise that these expressions also appear, after transforming to tangent bundle, in Lagrangian constraint analysis.

Since the analysis of a class of theories greatly benefits from the realisation of diffeomorphism invariance requirements, its validity relies on the validity of diffeomorphism invariance requirements. In particular, the analysis is valid as long as theories are invariant under $x^\m\mapsto x^\m - \e^\m(x)$ such that $\pa_i\e^0\neq 0$ (see Appendix \ref{app:diffeo}). This works for example if the theory is invariant under diffeomorphism transformation, in which $\e^\m$ is arbitrary.
If one puts a theory in a fixed background and demand that isometry of spacetime is preserved, then the analysis of our work would apply if the Killing vector $\e^\m$ satisfies $\pa_i\e^0\neq 0.$ So this works if the fixed background is, for example, flat spacetime or de Sitter spacetime. 

Secondary-constraint enforcing relations we have obtained in this paper is a correction to \cite{ErrastiDiez:2019trb}, \cite{ErrastiDiez:2019ttn}.
This means that behaviour of some theories are previously misjudged. We have shown in subsection \ref{subsec:samples} an example of a legitimate theory previously misinterpreted as containing extra degrees of freedom as well as an example of undesired theory with extra degrees of freedom previously misinterpreted as being legitimate. We leave the work of identifying or constructing all of the theories which pass the secondary-constraint enforcing relations and the completion relations for future. Nevertheless, a consequence can readily be discussed and is provided in subsection \ref{subsec:implications} which points out that legitimate terms previously misjudged could be reintroduced into models to investigate cosmological implications.

An important future work is to analyse a larger class of theories, not necessarily restricted to those describing only vector fields. In fact, an important step has already been laid out by \cite{ErrastiDiez:2020dux}, which gives criteria for counting the number of degrees of freedom for theories with Lagrangians as functions of up to first order derivative in fields. These criteria, however, should be revised because as points out by \cite{deRham:2023brw}, the analysis of \cite{ErrastiDiez:2020dux} is not correct even in the case of the standard Proca theory. Additionally, as reported in our paper, the analysis of \cite{ErrastiDiez:2020dux} when specialised to multi-field generalised Proca theories misses terms in intermediate steps, for example $\pa_i \ddot{Q}^\b$ and $\pa_i\pa_j \ddot{Q}^\b$ within $\dot\f_\a.$ The corrections are required to address these issues. Once they are taken care of, we expect that the analysis would benefit from the help of diffeomorphism conditions. 
This is because in constaint analysis, even for Lagrangian constraint analysis, time and space are not treated on an equal footing. So the manifestation of diffeomorphism invariance (or, in case of flat spacetime, Lorentz isometry) is lost in the steps. The manifestation could be recovered with the use of diffeomorphism invariance requirements.

\backmatter

\bmhead{Acknowledgments}

	We are grateful to Sheng-Lan Ko for interests and discussions. We would also like to thank Claudia de Rham for interests, discussions, and comments. S.J. is supported by the Research Professional 
Development Project Under the Science Achievement 
Scholarship of Thailand (SAST). 
\\
This preprint has not undergone peer review or any post-submission improvements or corrections. The Version of Record of this article is published in General Relativity and Gravitation, and is available online at https://doi.org/10.1007/s10714-023-03191-8.
\\
\\
\textbf{Data Availability} No data was analysed or generated in this work.\\
\\
\\
\textbf{Conflicts of interest} The authors declare that there is no conflict of interests.

\begin{appendices}

\section{Conditions from diffeomorphism invariance}\label{app:diffeo}
In this appendix, we consider a class of theories described in section \ref{sec:analysis}. Since these theories are diffeomorphism invariant, their Lagrangians would satisfy the conditions to be presented in this appendix.

Under diffeomorphism $x^\m\mapsto x^\m - \e^\m(x),$
the vector fields transform as
\be
\d_\e A_\m^\a
=\e^\n\pa_\n A_\m^\a + A_\n^\a\pa_\m\e^\n,
\ee
and the external fields $K$ transform under standard diffeomorphism.
The Lagrangian density transforms as
\be
\d_\e\cL
=\pa_\m(\e^\m\cL).
\ee

The only term containing
$\dot{A}_0^\a\dot{A}_0^\b$ in $\d_\e\cL - \pa_\m(\e^\m\cL)$
is
\be
\frac{\pa U_\b}{\pa\pa_i A_0^\a}\pa_i\e^0\dot A_0^\a\dot A_0^\b.
\ee
Demanding this expression to vanish gives
\be\label{diffeo-UpaiA0}
\frac{\pa U_\a}{\pa\pa_i A_0^\b}
+\frac{\pa U_\b}{\pa\pa_i A_0^\a}
=0.
\ee

Let us next turn to the coefficients of $\dot{A}_0^\b.$
We have

\be\label{diffeo-A0dot}
\begin{split}
	&\frac{\pa T}{\pa\pa_k A_0^\b}\pa_k\e^0
	+\frac{\pa T}{\pa\dot{A}_k^\b}\pa_k\e^0
	+\frac{\pa U_\b}{\pa A^\a_\n}A^\a_\m\pa_\n \e^\m
	+\frac{\pa U_\b}{\pa K}\d_\e K
	+2U_\b\dot{\e}^0
	-\pa_\m\e^\m U_\b
			\\
	&\quad
	-\e^\m \frac{\pa U_\b}{\pa K}\pa_\m K
	+\frac{\pa U_\b}{\pa\pa_i A^\a_\n}(\pa_i\e^\m\pa_\m A^\a_\n + \pa_iA^\a_\m\pa_\n \e^\m
	+A^\a_\m\pa_i\pa_\n \e^\m)\bigg|_{\dot{A}_0^\a = 0}
	=0.
\end{split}
\ee
It is worth noting that the precise form of $K$ are not important for subsequent calculations.
Taking derivative of eq.\eq{diffeo-A0dot} with respect to
$\dot A_j^\a$ gives
\be\label{cond-A0dot-1}
\begin{split}
	&\frac{\pa^2 T}{\pa\pa_k A_0^\b\pa \dot A_j^\a}
	+\frac{\pa^2 T}{\pa\dot{A}_k^\b\pa \dot A_j^\a}
	+\frac{\pa U_\b}{\pa\pa_k A_j^\a}
	=0.
\end{split}
\ee
Let us take derivative of eq.\eq{diffeo-A0dot} with respect to
$\pa_j A_0^\a$, then swap the indices $\a$ and $\b$, add it to the original equation, and use eq.\eq{diffeo-UpaiA0}, we obtain
\be\label{cond-A0dot-2}
\begin{split}
	&2\frac{\pa^2 T}{\pa\pa_j A_0^{(\a}\pa\pa_k A_0^{\b)}}
	\!+\!2\frac{\pa^2 T}{\pa\pa_j A_0^{(\a}\pa\dot{A}_k^{\b)}}
	\!+\!\frac{\pa U_{\b}}{\pa\pa_j A^{\a}_k}
	\!+\!\frac{\pa U_{\a}}{\pa\pa_j A^{\b}_k}
	=0.
\end{split}
\ee

Expressing in phase space, the conditions eq.\eq{cond-A0dot-1}-\eq{cond-A0dot-2} become
\be\label{cond-A0dot-phasespace-1}
\begin{split}
	&\frac{\pa^2 \cT}{\pa\pa_k A_0^\b\pa \L_j^\a}
	+\frac{\pa^2 \cT}{\pa\L_k^\b\pa \L_j^\a}
	+\frac{\pa U_\b}{\pa\pa_k A_j^\a}
	=0.
\end{split}
\ee
\be\label{cond-A0dot-phasespace-2}
\begin{split}
	&2\frac{\pa^2 \cT}{\pa\pa_j A_0^{(\a}\pa\pa_k A_0^{\b)}}
	\!+\!2\frac{\pa^2 \cT}{\pa\pa_j A_0^{(\a}\pa\L_k^{\b)}}
	\!+\!\frac{\pa U_{\b}}{\pa\pa_j A^{\a}_k}
	\!+\!\frac{\pa U_{\a}}{\pa\pa_j A^{\b}_k}
	=0.
\end{split}
\ee
By substituting eq.\eq{cond-A0dot-phasespace-1}
into eq.\eq{cond-A0dot-phasespace-2},
we obtain
\be\label{cond-A0dot-phasespace-3}
\frac{\pa^2 \cT}{\pa\pa_{(j|} A_{0}^\a\pa\pa_{|k)} A_0^\b}
-\frac{\pa^2 \cT}{\pa\L_{(j}^\a\pa\L_{k)}^\b}
=0.
\ee

It can easily be seen that the analysis in this appendix is valid as long as $\pa_i\e^0\neq 0.$
In particular, it is valid when the theories are diffeomorphism invariant. It is also valid when the theory is put in flat spacetime, in which the Lorentz isometry, with Killing vector $\e^\m = \o^\m{}_\n x^\n$ for $\o_{\m\n} = -\o_{\n\m}$, is required.

\section{\label{app:paphipaQdot}Expressions of $\pa\f_\a/\pa\pa_{\cI}\dot{Q}^\b$ in phase space}

In this appendix, we outline necessary steps to express
$\pa\f_\a/\pa\pa_{\cI}\dot{Q}^\b$ in phase space.
We use the same set-up and notations as those given in sections \ref{sec:analysis}-\ref{sec:checks}. For convenient, let us denote $P_A$ and $\L^A$ as collective for $\p_a^i$ and $\L^\a_i,$ respectively.

The idea is to first express $\pa\f_\a/\pa\pa_{\cI}\dot{Q}^\b$ in terms of $\a_M$. This gives
\be\label{phipaq}
\frac{\pa\f_\a}{\pa\pa_{\cI}\dot{Q}^\b}
=\frac{\pa\a_\a}{\pa\pa_{\cI}Q^\b}
-\frac{\pa\a_\a}{\pa\pa_{\cJ}\dot{Q}^A}\frac{\pa\pa_{\cJ}(M^{AB}\a_B)}{\pa\pa_{\cI}\dot{Q}^\b}.
\ee
Next, we directly express $\a_A$ in terms of Lagrangian then transforming to phase space, but transform $\a_\a$ to $-\tilde{\Omega}_\a$ (cf. eq.\eq{Oaaa}). This gives
\be\label{mpapaq}
\frac{\pa\pa_k(M^{AB}\a_B)}{\pa\pa_i\pa_j\dot{Q}^\b}
=M^{AB}\frac{\pa\a_B}{\pa\pa_l\dot{Q}^\b}\d_{(k}^i\d_{l)}^j,
\ee
\be
\frac{\pa\pa_j(M^{AB}\a_B)}{\pa\pa_i\dot{Q}^\b}
=\pa_j\lrbrk{M^{AB}\frac{\pa\a_B}{\pa\pa_i\dot{Q}^\b}}+\d_j^iM^{AB}\frac{\pa\a_B}{\pa\dot{Q}^\b},
\ee
\be
\frac{\pa\pa_j(M^{AB}\a_B)}{\pa\dot{Q}^\b}=\pa_j\lrbrk{M^{AB}\frac{\pa\a_B}{{\pa\dot{Q}^\b}}},
\ee
\be
\frac{\pa\a_B}{\pa\pa_i\dot{Q}^\b}=\frac{\pa^2 \cT}{\pa\pa_iQ^\b\pa\L^B}
+\frac{\pa U_\b}{\pa \pa_iQ^B},
\ee
\be\label{aq-pu}
\frac{\pa\a_B}{\pa\dot{Q}^\b}
=\frac{\pa^2\cT}{\pa Q^\b\pa\L^B}
+\pa_i\lrbrk{\frac{\pa U_\b}{\pa\pa_iQ^B}}
-\frac{\pa U_\b}{\pa Q^B},
\ee
\be
\frac{\pa\a_\a}{\pa\pa_i\pa_jQ^\b}
=-\frac{\pa\ot_\a}{\pa\pa_i\pa_jQ^\b}
-\frac{\pa \ot_\a}{\pa\pa_{(i|}P_B}\frac{\pa^2\cT }{\pa\pa_{|j)}Q^\b\pa\L^B},
\ee
\be
\begin{split}
	\frac{\pa\a_\a}{\pa\pa_iQ^\b}
	&=-\frac{\pa\ot_\a}{\pa\pa_iQ^\b}
	-\frac{\pa\ot_\a}{\pa\pa_\cI P_B}\pa_\cI\lrbrk{\frac{\pa^2\cT }{\pa\pa_iQ^\b\pa\L^B}}
	-\frac{\pa\ot_\a}{\pa\pa_iP_B}\frac{\pa^2\cT}{\pa Q^\b\pa\L^B},
\end{split}
\ee
\be
\frac{\pa\a_\a}{\pa Q^\b}
=-\frac{\pa\ot_\a}{\pa Q^\b}
-\frac{\pa\ot_\a}{\pa\pa_\cI P_B}\pa_\cI\lrbrk{\frac{\pa^2\cT }{\pa Q^\b\pa\L^B}},
\ee
\be
\frac{\pa\a_\a}{\pa\pa_i\dot{Q}^A}
=-\frac{\pa\ot_\a}{\pa\pa_iP_B}W_{AB},
\ee
\be\label{aq-o}
\frac{\pa\a_\a}{\pa\dot{Q}^A}
=-\frac{\pa\ot_\a}{\pa\pa_iP_B}\pa_iW_{BA}-\frac{\pa\ot_\a}{\pa P_B}W_{BA}.
\ee
By substituting eq.\eq{mpapaq} - \eq{aq-o} into eq.\eq{phipaq}, we obtain
\be\label{f0-f1-f2}
\begin{split}
	\frac{\pa\f_\a}{\pa\dot{Q}^\b}
	&=-\cC_{0\b\a}
	+\pa_i\cC_{1\b\a}^i
	-\pa_i\pa_j\cC_{2\b\a}^{ij},\\
	\frac{\pa\f_\a}{\pa\pa_i\dot{Q}^\b}
	&=\cC_{1\b\a}^i
	-2\pa_j\cC_{2\b\a}^{ij},\\
	\frac{\pa\f_\a}{\pa\pa_i\pa_j\dot{Q}^\b}
	&=-\cC_{2\b\a}^{ij}.
\end{split}
\ee
By using diffeomorphism invariance requirements,  eq.\eq{C1-C2-properties} is realised. This simplifies eq.\eq{f0-f1-f2}. Further simplifications are possible.
For this, let us note that using eq.\eq{phipaq}-\eq{aq-pu} and diffeomorphism invariance requirements, one obtains
\be
\begin{split}
	&\frac{\pa\f_\a}{\pa\dot{Q}^\b}
	-\frac{\pa\f_\b}{\pa\dot{Q}^\a}
	+\pa_i\lrbrk{\frac{\pa\f_\b}{\pa\pa_i\dot{Q}^\a}}
	=\frac{\pa\a_\a}{\pa Q^\b}
	-\frac{\pa\a_\b}{\pa Q^\a}
	+\pa_i\lrbrk{\frac{\pa \a_\b}{\pa\pa_iQ^\a}
		+\frac{\pa\a_\b}{\pa\dot{Q}^\a_i}
		+\frac{\pa\a_\a^i}{\pa\dot{Q}^\b}}.
\end{split}
\ee
Then by expressing $\a_M$ in terms of Lagrangian and using diffeomorphism invariance and secondary-constraint enforcing relations, we obtain
\be\label{c0=c0-pac1-equl}
\frac{\pa\f_\a}{\pa\dot{Q}^\b}
-\frac{\pa \f_\b}{\pa\dot{Q}^\a}
+\pa_i\lrbrk{\frac{\pa\f_\b}{\pa\pa_i\dot{Q}^\a}}
=0,
\ee
which is equivalent to the phase space expression
\be\label{c0=c0-pac1}
\cC_{0\a\b}
=\cC_{0\b\a}-\pa_i\cC_{1\b\a}^i.
\ee
Finally, this gives
\be\label{f0-f1-f2-final}
\frac{\pa\f_\a}{\pa\dot Q^\b}
= -\cC_{0\a\b},\quad
\frac{\pa\f_\a}{\pa\pa_i\dot Q^\b}
=-\cC^{i}_{1\a\b},\quad
\frac{\pa\f_\a}{\pa\pa_i\pa_j\dot Q^\b}
=0.
\ee

\end{appendices}

\end{document}